\documentclass[journal,10pt,twocolumn]{IEEEtran}

\usepackage[cmex10]{amsmath}
\usepackage{amssymb}
\usepackage{amsthm}
\usepackage{graphicx}
\usepackage{color}
\usepackage{bm}
\usepackage{bbm}
\usepackage{url}
\usepackage{array}
\usepackage{multicol}
\usepackage{subfigure}

\usepackage{algorithm}
\usepackage{algpseudocode}

\usepackage[colorlinks]{hyperref}
\usepackage{url}

\theoremstyle{remark}

\usepackage{array}
\makeatletter
\newcommand{\thickhline}{%
    \noalign {\ifnum 0=`}\fi \hrule height 1pt
    \futurelet \reserved@a \@xhline
}
\newcolumntype{"}{@{\hskip\tabcolsep\vrule width 1pt\hskip\tabcolsep}}
\makeatother

\begin{document}

\title{Diffusion Fluid Antenna Systems for Resilient ISAC}

\author{\IEEEauthorblockN{Noor Waqar, 
		                           Kai-Kit Wong, \emph{Fellow, IEEE},
		                           Chan-Byoung Chae, \emph{Fellow, IEEE}, and 
		                           Ross Murch, \emph{Fellow, IEEE}
}
\vspace{-8mm}


\thanks{The work of K. K. Wong is supported by the Engineering and Physical Sciences Research Council (EPSRC) under grant EP/W026813/1.}
\thanks{The work of C.-B. Chae was supported by the Korean Government under Grant IITP-2025-RS-2024-00428780 and Grant IITP-RS-2026-25489110.}
\thanks{The work of R. Murch was supported by the Hong Kong Research Grants Council Area of Excellence Grant AoE/E-601/22-R.}

\thanks{N. Waqar and K. K. Wong are with the Department of Electronic and Electrical Engineering, University College London, London WC1E 7JE, United Kingdom. K. K. Wong is also affiliated with Yonsei Frontier Laboratory, Yonsei University, Seoul, 03722, Republic of Korea.}
\thanks{C. B. Chae is with School of Integrated Technology, Yonsei University, Seoul, 03722, Republic of Korea.}
\thanks{R. Murch is with the Department of Electronic and Computer Engineering, Hong Kong University of Science and Technology, Clear Water Bay, Hong Kong SAR, China.}

\thanks{Corresponding author: Kai-Kit Wong (e-mail: $\rm kai\text{-}kit.wong@ucl.ac.uk$).}
}
\maketitle

\begin{abstract}
Most existing integrated sensing and communication (ISAC) research focuses on enabling a base station (BS) to support sensing and communication over shared resources through advanced waveform design and power allocation. Yet, the object-side perspective remains underexplored. For example, it may be required to make an object difficult to detect for security reasons. Moreover, an object in close proximity to a target may generate dominant reflections that confuse the BS and impair sensing reliability for the intended target. These challenges motivate the fluid antenna system (FAS) paradigm which introduces a reconfigurable spatial degree of freedom (DoF) that can reshape object sensing signatures via port selection, beyond what waveform and power control alone can provide. This paper proposes diffusion FAS, a generative artificial intelligence (AI)-driven framework that takes advantage of spatial agility to steer ISAC operation over the electromagnetic fading manifold. Instead of optimizing ISAC solely in the power domain, diffusion FAS casts ISAC as a \emph{dynamic spatial selection} problem in which antenna states (i.e., ports) are selected to shape sensing signatures while maintaining communication objectives. To handle sparse measurements, we will employ a conditional denoising diffusion probabilistic model (DDPM) to reconstruct the latent spatial correlation structure from a small set of observed ports, enabling efficient exploration of the reconfigurable aperture. We demonstrate two FAS-enabled ISAC modes: (1) \emph{generative spatial stealth}, which identifies localized deep fades to suppress a target's sensing visibility by up to two orders of magnitude, and (2) \emph{target isolation}, which creates spatial nulls that reject interference from adjacent objects. 
Our results show that combining generative priors with reconfigurable apertures offers a viable solution for secure and resilient \emph{ISAC}.
\end{abstract}


\vspace{-2mm}
\section{Introduction}
\subsection{Background}
\IEEEPARstart{T}{he pursuit} toward sixth-generation (6G) wireless networks heralds an unprecedented unification of extreme-high rate access, super-massive connectivity, and ultra-reliable communication under stringent latency constraints \cite{Tariq-2020,wang2023road6g,jiang2021road6g}. A further recent trend has seen 6G discussion dominated by integrated sensing and communication (ISAC), a paradigm shift that recognizes the rising importance of sensing beyond the primary communication-oriented mobile networks \cite{Liu-2022jsac,Wu-2025jsteap,Ghosh-2025jsteap}. To support high-precision sensing while meeting stringent communication requirements, a substantial body of work has been reported \cite{Liu-2022comst,Zhang-2022comst}, which focuses on developing advanced waveform and power allocation to enable a base station (BS) to support both sensing and communication tasks over shared resources. Moreover, multiple-input multiple-output (MIMO) architectures play a central role in enabling ISAC by offering the spatial degrees of freedom (DoF) needed to jointly shape communication and sensing signals \cite{Zhang-mimoisac2024}.

The benefits of MIMO are well established, and the evolution toward extra-large MIMO appears compelling \cite{wang2024XLMIMO}. Yet, its scalability limitations are increasingly evident. First, even in communications without sensing requirements, fifth generation (5G) New Radio (NR) precoding is far from trivial: it relies on the maximization of signal-to-interference plus noise ratio (SINR) based on quantized channel state information (CSI), which imposes substantial feedback and overhead \cite{Villalonga2022spectral}. More worryingly, the increase in the peak-to-average power ratio as a result of precoding greatly degrades the efficiency of power amplifiers at each antenna, causing the power consumption to explode at the BS when the antenna count is very large \cite{hung2014papr_mimo_ofdm}. These issues are expected to worsen under ISAC settings.

On the other hand, despite the exciting prospect of ISAC, it raises security and privacy concerns \cite{Wei-2022cmag}. 

\begin{quote}
\centering
{\em What if we require an object to be hidden?\\
How can we suppress an object's reflections and reduce its sensing visibility?} 
\end{quote}
To respond to these concerns, the authors of \cite{Qu-2024iot} introduced a security and privacy-preserving network (SPPN) for information handling and collaborative efforts among trusted parties. In the physical layer, jamming was considered. In \cite{Ma-2025rwc}, these problems were tackled using advanced strategies such as null-space projection and the use of artificial noise. Recently, \cite{Zhu-2025comst} provided an architecture defined into three layers: the hardware layer, the omniscient layer, and the application layer. The idea is to consider new design principles for artificial intelligence (AI)-empowered networks that support security in ISAC. 

Although AI can help a receiver \emph{detect} and \emph{interpret} sensing outcomes, it has limited leverage in addressing object-side objectives such as \emph{reducing} sensing visibility. In particular, once an external transmitter illuminates the environment, the object's reflections are governed primarily by electromagnetic scattering physics and the geometry/material properties of the object, which cannot be altered at the sensing node. As a result, purely AI-driven approaches are confined to post-processing, e.g., improving detection, classification, or interference suppression, rather than actively preventing an object from being sensed. Sending a jamming signal from the sensing object can be effective in reducing its sensing visibility but the use of jamming is controversial as it preemptively pollutes the entire radio environment. Achieving true sensing suppression hence  requires additional controllable DoFs at the physical layer. In this regard, the fluid antenna system (FAS) that considers the utilization of reconfigurable apertures capable of reshaping the propagation/scattering process, can be the solution.

\vspace{-2mm}
\subsection{The FAS Paradigm}
First investigated in \cite{wong2020_fas_limits,wong2021_fas_twc}, FAS is a hardware-agnostic system concept that considers the antenna as a reconfigurable physical-layer resource to broaden system design and network optimization \cite{new2025_fas_tutorial,hong2026_fas_survey,new2026_fas_jsac,lu2025_fluid_antennas_mag,FAS_wu_tuo1}. In practice, it can be implemented using a range of antenna technologies such as movable elements \cite{zhu2024_fas_history}, liquid antennas \cite{shen2024_surfacewave_fas,wang2026_em_reconfig_fas}, metamaterials \cite{Zhang-jsac2026,liu2025_meta_fluid_optics}, reconfigurable pixels \cite{zhang2025_pixel_reconfig,liu2025_wideband_pixel_fas,wong2026_pixel_meet_fas}. The theoretical premise is profound: even within a relatively small physical footprint, the multipath fading observed across FAS positions exhibit rich spatial fluctuations, and FAS yields extraordinary diversity and reliability gains \cite{khammassi2023_analytical_fas,new2024_fas_outage, ramirez2024_blockcorr,psomas2024_diversity_coded,vega2024_simple_nakagami,vega2024_outage_diversity,alvim2024_alpha_mu,ghadi2023_copula_cl,ghadi2024_gaussian_copula,new2024_mimofas_infotheory}. CSI estimation for FAS has been recently addressed in \cite{xu2024_channel_est_mmwave_fas,zhang2024_successive_bayes_fas,xu2025_sbl_fas}.

Beyond point-to-point single-user links, this spatial diversity has also catalyzed the development of \emph{fluid antenna multiple access} (FAMA), where port selection serves as a fundamental multiple-access dimension \cite{wong2022_fama,wong2023_fast_fama,wong2023_slow_fama, xu2024_two_user_outage_fama,wong2024cuma,wong2023_opportunistic_fama}. In recent years, channel coding \cite{hong2025coded,hong20255gcoded,hong2025downlink} and learning-based methods \cite{waqar2026_turbocharging_fama, waqar2026_attentional_copula_fama,waqar2023_dl_slow_fama,eskandari2024_cgan_slow_fama,waqar2024_opportunistic_fama_rl} have led to powerful FAMA schemes, capable of massive connectivity without BS precoding/optimization. A contemporary survey on FAMA can be found in \cite{fama-overview2026}. In a broader sense, AI and large language models (LLMs) also have been demonstrated to be well suited for optimizing FAS in many applications \cite{wang2024_ai_empowered_fas,wang2025largex,Wang-2026wcl}.

In terms of security, FAS as an additional spatial DoF has been shown to significantly enhance secrecy performance \cite{tang2023_secret_fas,xu2024_jamming_fas,ghadi2024_security_fas,vega2024_secrecy_outage,ghadi2025_covert_fas}. On the other hand, ISAC is also becoming more achievable when FAS is employed \cite{wang2024_fas_isac_drl, zhou2024_fas_isac_wcl,zou2024_isac_tradeoff,meng2025_isac_network}. However, whether and how FAS can be utilized for securing ISAC is not well understood.

\vspace{-2mm}
\subsection{From Target Detection to Physical-Layer Stealth}
\begin{figure*}[]
\centering
\subfigure[]{\includegraphics[width=.4\linewidth]{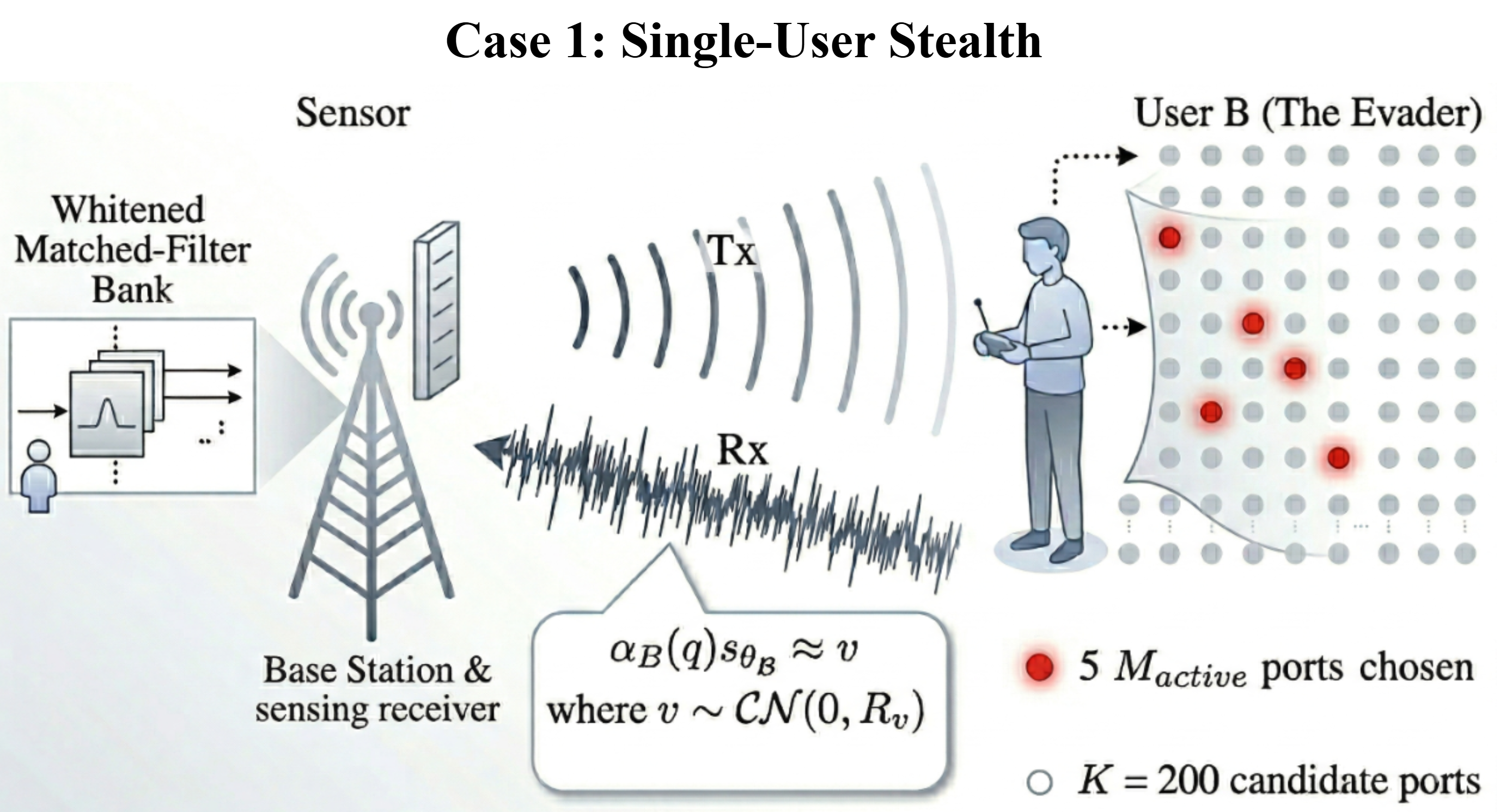}\label{fig:case1}}
\hspace{5mm}
\subfigure[]{\includegraphics[width=0.38\linewidth]{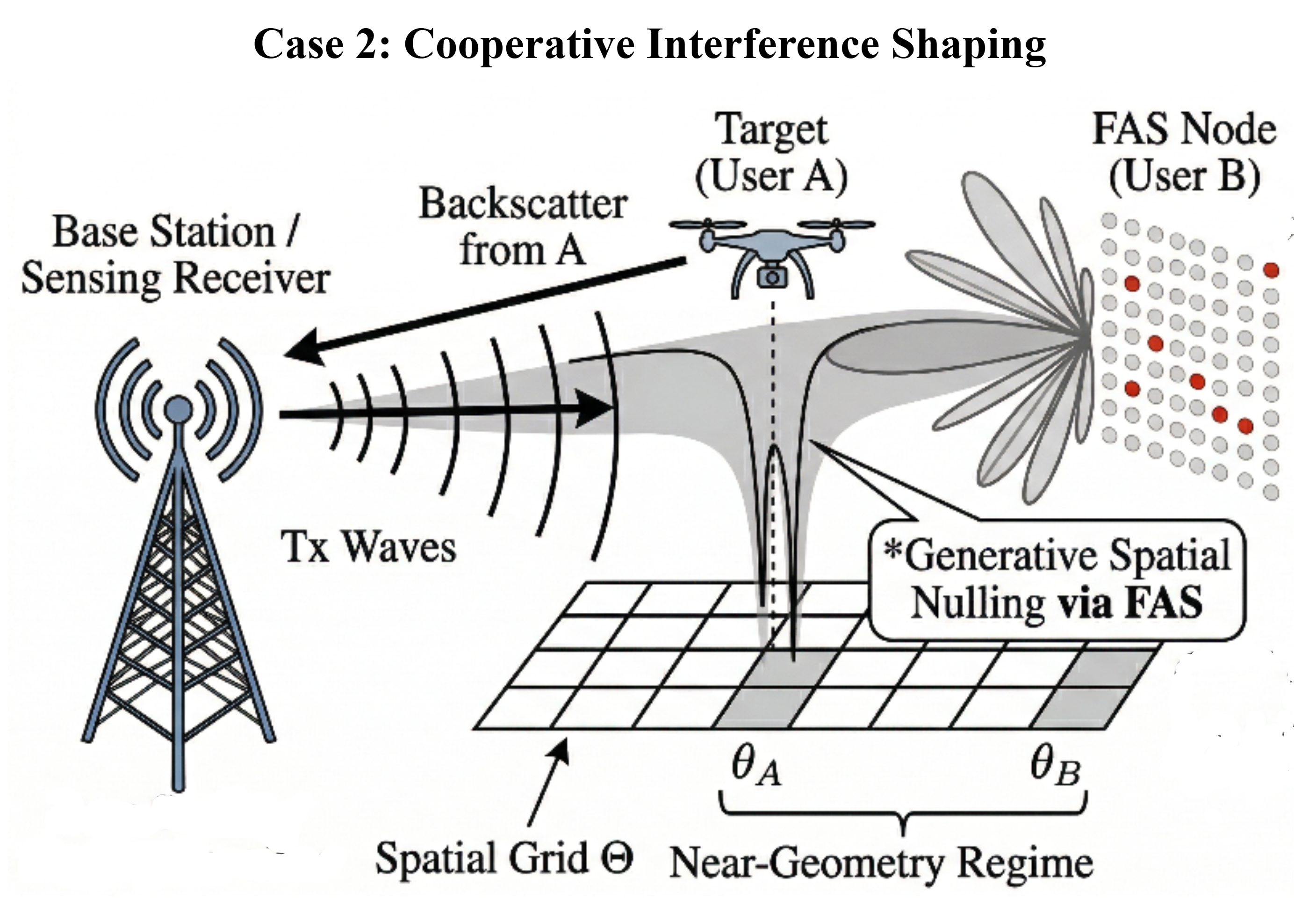}\label{fig:case2}}
\caption{ISAC with an FAS-enabled object (i.e., User B): (a) Use Case I (single-user stealth); and (b) Use Case II (cooperative interference shaping).}\label{fig:secureISAC}
\vspace{-4mm}
\end{figure*}

If equipped with FAS, a reconfigurable object is no longer merely a passive scatterer. By changing its activated ports, it alters its effective complex coupling coefficient (and equivalently, its radar cross-section) as observed by a distant sensing receiver. This leads to two highly unique sensing-centric use cases that deviate entirely from standard ISAC literature:
\begin{enumerate}
\item \textbf{Use Case I (Single-User Stealth)}---A FAS-enabled user actively seeks to evade detection by a sensing receiver operating a matched-filter bank. Instead of relying on complex waveform jamming, the user dynamically adjusts its multi-port selection to minimize its effective coupling magnitude, forcing its backscattered return to statistically emulate colored background clutter. This conceptually advances recent physical-layer security and covert communication perspectives in FAS, framing port selection as an anti-sensing mechanism, see Fig.~\ref{fig:case1}.
\item \textbf{Use Case II (Cooperative Interference Shaping)}---A sensing receiver aims to detect and localize a specific Target-of-Interest. Unfortunately, another user is located in the \emph{near-geometry} regime (i.e., occupying adjacent angular or delay bins), and the backscatter from this user creates severe main-lobe interference, inflating the constant false alarm rate (CFAR) and significantly degrading the localization accuracy of the actual target. In our framework, this user is equipped with FAS which intelligently selects a sparse port mask designed to place a deep spatial null explicitly in the sensing receiver's \emph{guard bins} surrounding the target, see Fig.~\ref{fig:case2}. 
\end{enumerate}

\vspace{-2mm}
\subsection{AI-Native Solutions}
Delivering the two use cases is challenging. FAS control is inherently \emph{combinatorial}; producing a binary length-$K$ mask under a strict cardinality constraint $M$ per snapshot, where $K$ denotes the number of available ports and $M$ is the number of radio-frequency (RF) chains, is NP-hard. Furthermore, the user terminal only possesses partial observability of the port-domain coupling vector. Exhaustive search is mathematically impossible as well within the coherence time of the channel. To circumvent these constraints, recent FAS research has increasingly leveraged AI, utilizing deep neural networks and generative adversarial networks (GANs) for port scheduling. However, traditional deep learning struggles to adapt to dynamic cardinality constraints without extensive retraining, and standard architectures fail to capture the highly structured, non-Gaussian nature of spatial interference fields. 

Consequently, we resort to denoising diffusion probabilistic models (DDPMs). Diffusion models have emerged as a powerful paradigm for learning structured physical priors and generating high-quality solutions under complex constraints, demonstrating remarkable success in wireless channel estimation and signal detection \cite{letafati2023_diffusion_wireless,wu2024_cddm,fesl2024_diffusion_mimo_ce, jin2024_gdm4mmimo,du2024_diffusion_network_opt,wang2025_diffusion_detection}. 

\vspace{-2mm}
\subsection{Contributions}
This paper proposes a radically new approach, referred to as a \emph{diffusion-controlled FAS policy}. Rather than treating port selection as a discrete classification problem, our generative framework operates in a continuous logit space. By using a differentiable soft-top-$M$ relaxation, we generate high-quality port-activation candidates that are iteratively \emph{guided} during the reverse diffusion process. This is done using a differentiable proxy energy function which directly calculates the sensing receiver's overlap kernel, allowing the generative model to steer the port selection toward optimal stealth or cooperative interference nulling. Our contributions are summarized below:
\begin{itemize}
\item We formulate a sensing snapshot model operating over a whitened matched-filter bank with colored clutter. Breaking from standard ISAC literature, we evaluate the performance of the FAS-controlled effective coupling using strict CFAR-calibrated metrics, focusing on the conditional localization accuracy, $\mathbb{P}(\mathrm{correct}|\mathrm{det})$, to prove that FAS protects the localization grid of adjacent targets.
\item We mathematically define and solve for two highly novel operational paradigms: (i) \emph{clutter-likeness/stealth} by intelligent port selection to evade sensing, and (ii) \emph{cooperative interference shaping} to actively carve out spatial nulls in the guard bins of a target-of-interest.
\item We design a conditional DDPM that circumvents the NP-hard discrete mask selection problem. By diffusing over continuous logits and applying a differentiable energy metric derived from the matched-filter overlap kernel, our AI policy respects strict cardinality budgets while minimizing sensing interference, extending recent diffusion-for-wireless advances into discrete FAS control.
\item We present numerical results to assess the FAS approach under varying hardware constraints \emph{without} retraining. Specifically, we sweep the number of active ports, the available spatial aperture, and the measurement budget (i.e., the number of probed antenna states per coherence block), and quantify the resulting detection and estimation performance. The results are provided for both finite-scattering and rich-scattering channel conditions.
\end{itemize}


\section{System Model}\label{sec:system_model}
We consider a sensing-centric coexistence scenario wherein a sensing receiver aims to detect and localize targets within a single coherent processing interval (snapshot) over a narrowband channel. Operating within this environment is a FAS-enabled node (User~B) whose multi-port selection is governed by a generative diffusion policy. Our unified signal model encapsulates two distinct operational paradigms:
\begin{itemize}
\item \textbf{Use Case I (Stealth/Clutter-Likeness)}---The diffusion-controlled FAS node operates in isolation and actively optimizes its port activation to minimize its effective coupling. The objective is to become statistically indistinguishable from background clutter, thereby minimizing its detectability at the sensing receiver.
\item \textbf{Use Case II (Cooperative Interference Shaping)}---Two uncoordinated nodes are present simultaneously. \emph{User~A} acts as the target-of-interest whose presence and spatial location are to be estimated. \emph{User~B} (i.e., the FAS-enabled node) leverages its spatial DoFs to dynamically shape its effective coupling, deliberately carving out a spatial null to improve User~A's detection and localization under a fixed CFAR threshold.
\end{itemize}

\vspace{-2mm}
\subsection{FAS Geometry and Port-Domain Channel Models}\label{subsec:fas_geometry_channel}
User~B is equipped with FAS providing $K$ candidate ports distributed within a finite physical aperture. Let $D \in \{1,2\}$ represent the dimensionality of the FAS, defining a set of candidate port locations $\mathcal{P} \triangleq \big\{\mathbf{p}_k \in \mathbb{R}^{D}: k=1,\dots,K\big\}$. 

Let $\mathbf{g}_B \in \mathbb{C}^{K}$ denote the \emph{port-domain coupling vector} (the baseband channel response) between User~B's $K$ candidate ports and the sensing receiver during the given snapshot:
\begin{equation}\label{eq:gB_def}
\mathbf{g}_B \triangleq [g_{B,1}, g_{B,2}, \dots, g_{B,K}]^{\mathsf{T}}.
\end{equation}
We model this coupling using a correlated Rayleigh fading framework, $\mathbf{g}_B \sim \mathcal{CN}\!\big(\mathbf{0},\,\sigma_{g}^{2}\mathbf{R}_K\big)$, where $\sigma_g^2$ is the average coupling strength and $\mathbf{R}_K \in \mathbb{C}^{K\times K}$ is the spatial correlation matrix induced by the geometry and the underlying scattering environment. We consider two canonical scattering regimes.

{\em Rich isotropic scattering}---In environments characterized by an infinitely many uniformly distributed scatterers, the spatial correlation is dictated strictly by port separation. The $(k,\ell)$-th element of $\mathbf{R}_K$ is given by
\begin{equation}\label{eq:rich_corr}
[\mathbf{R}_K]_{k,\ell}
= J_0\!\left(\frac{2\pi}{\lambda}\|\mathbf{p}_k-\mathbf{p}_\ell\|_2\right),
\end{equation}
where $J_0(\cdot)$ denotes the zeroth-order Bessel function of the first kind, and $\lambda$ denotes the carrier wavelength.

{\em Finite scattering}---To capture environments dominated by a limited number of scattering clusters (e.g., millimeter-wave channels), we adopt a finite-scattering model with $L$ dominant paths. The spatial correlation is formulated as
\begin{equation}\label{eq:finite_corr}
[\mathbf{R}_K]_{k,\ell}
= \frac{1}{L}\sum_{i=1}^{L}
\cos\!\left(\frac{2\pi}{\lambda}\mathbf{u}_i^{\mathsf{T}}(\mathbf{p}_k-\mathbf{p}_\ell)\right),
\end{equation}
in which $\{\mathbf{u}_i\}_{i=1}^{L}$ are unit direction vectors corresponding to the angle-of-departure (AoD) or angle-of-arrival (AoA) of the $i$-th multipath component. Smaller values of $L$ yield a highly structured, low-rank correlation matrix, whereas $L \to \infty$ asymptotically approaches the rich scattering model.

\vspace{-2mm}
\subsection{ISAC Signal Model and Effective FAS Coupling}\label{subsec:isac_signal_model}
The sensing receiver attempts to localize a target over a discrete spatial/delay grid $\Theta \triangleq \mathbb{Z}_{N_\theta} = \{0,1,\dots,N_\theta-1\}$. It employs a known template dictionary evaluated across the grid, $\mathbf{S} \triangleq \big[\mathbf{s}_0,\,\dots,\,\mathbf{s}_{N_\theta-1}\big] \in \mathbb{C}^{d\times N_\theta}$, where $\mathbf{s}_\theta \in \mathbb{C}^{d}$ is the unit-norm signature vector ($\|\mathbf{s}_\theta\|_2^2 = 1$) associated with grid point $\theta \in \Theta$, and $d$ is the effective feature dimension.

Within any given snapshot, the BS receiver observes a $d$-dimensional complex feature vector comprising the superposition of the target (User A) and the FAS node (User B):
\begin{equation}\label{eq:rx_model}
\mathbf{x}=\alpha_A \mathbf{s}_{\theta_A}+\alpha_B(\mathbf{q}) \mathbf{s}_{\theta_B}+\mathbf{v},
\end{equation}
in which $\theta_A, \theta_B \in \Theta$ are the respective grid locations, and $\alpha_A \in \mathbb{C}$ denotes the unknown scattering/echo coefficient of the target. Also, the term $\mathbf{v}$ models the aggregation of colored clutter and thermal noise, distributed as $\mathbf{v} \sim \mathcal{CN}(\mathbf{0},\mathbf{R}_v)$, where $\mathbf{R}_v \triangleq \sigma_c^2 \mathbf{R}_c + \sigma_n^2 \mathbf{I}_d$, and $\mathbf{R}_c$ represents the normalized spatial covariance matrix of the clutter.

{\em FAS state-coding and effective coupling}---Instead of traditional port selection where inactive ports electromagnetically ``disappear,'' we adopt a physically rigorous load-modulated scattering model. Specifically, an activated port represents an active reflection state (State 1) while an inactive port represents a baseline reflection state (State 0). Let $\rho_0$ and $\rho_1$ denote the complex reflection coefficients associated with State 0 and State 1, respectively. User~B controls its footprint via a binary state vector $\mathbf{q} \in \{0,1\}^K$, setting exactly $M_{\mathrm{active}}$ ports to State 1 and the remaining $K - M_{\mathrm{active}}$ ports to State 0. The aggregate port reflection vector is given by
\begin{equation}\label{eq:port-reflect-vector}
\boldsymbol{\rho}(\mathbf{q}) \triangleq \rho_0\mathbf{1} + (\rho_1-\rho_0)\mathbf{q} \in \mathbb{C}^{K}.
\end{equation}

Thus, User~B's \emph{effective complex coupling} is defined as the spatial inner product of the channel and the reflection vector, scaled by its baseline scattering coefficient $\alpha_B^{(0)}$ and normalized by $\sqrt{K}$ to stabilize the average interference variance
\begin{equation}\label{eq:effective_alphaB}
\alpha_B(\mathbf{q}) \triangleq \frac{\alpha_B^{(0)}}{\sqrt{K}} \mathbf{g}_B^{\mathsf{H}} \boldsymbol{\rho}(\mathbf{q}),
\end{equation}
which establishes the fundamental link between the physical FAS hardware and the sensing model. If $\mathbf{q} = \mathbf{0}$, the node naturally scatters energy ($\boldsymbol{\rho}(\mathbf{0}) = \rho_0\mathbf{1}$). The combinatorial configuration of $\mathbf{q}$ acts as a spatial codeword that manipulates the amplitude and phase of $\alpha_B(\mathbf{q})$, modulating the rank-one interference term at the sensing receiver, see (\ref{eq:rx_model}).

{\em Interference geometry and coupling}---User~B's presence critically degrades the sensing accuracy of User~A when $\theta_B$ is close to $\theta_A$. To capture this \emph{near-geometry} regime, we model the locations as coupled: $\theta_B = (\theta_A + \Delta)\bmod N_\theta$, where $\Delta$ denotes a localized spatial offset set in grid bins.

\vspace{-2mm}
\subsection{Partial Observability Constraints}\label{subsec:partial_observability}
User~B cannot perfectly estimate the full $K$-dimensional coupling vector $\mathbf{g}_B$ within a single symbol duration. Let $\mathbf{m} \in \{0,1\}^{K}$ be an observation mask satisfying $\mathbf{1}^{\mathsf{T}}\mathbf{m} = M_{\mathrm{obs}}$, where $M_{\mathrm{obs}} \ll K$. Consequently, User B observes only a sparse, noisy subset of the channel, $\tilde{\mathbf{g}}_B = \mathbf{m} \odot \mathbf{g}_B$. To facilitate the subsequent AI-driven policy, the available instantaneous knowledge is encoded into a real-valued observation vector:
\begin{equation}\label{eq:policy_obs}
\mathbf{o}_B \triangleq \big[ \mathbf{m}^{\mathsf{T}}, \, \Re\{\tilde{\mathbf{g}}_B\}^{\mathsf{T}}, \, \Im\{\tilde{\mathbf{g}}_B\}^{\mathsf{T}} \big]^{\mathsf{T}} \in \mathbb{R}^{3K}.
\end{equation}

\vspace{-2mm}
\subsection{Whitened Matched-Filter Detection and Metrics}\label{subsec:detection_localization}
To localize targets, the sensing receiver employs a whitened matched-filter bank. The test statistic at each candidate grid point $\theta \in \Theta$ is the normalized whitened projection:
\begin{equation}\label{eq:T_theta}
T(\theta;\mathbf{x})
\;\triangleq\;
\frac{\big|\mathbf{s}_\theta^{\mathsf{H}}\mathbf{R}_v^{-1}\mathbf{x}\big|^2}
{\mathbf{s}_\theta^{\mathsf{H}}\mathbf{R}_v^{-1}\mathbf{s}_\theta}.
\end{equation}
The global detector and the corresponding coordinate estimate are given by the global maximization over the grid:
\begin{equation}\label{eq:det_and_hat}
\left\{\begin{aligned}
T_{\max}(\mathbf{x})& \triangleq \max_{\theta\in\Theta}T(\theta;\mathbf{x}),\\
\hat{\theta} &\triangleq \arg\max_{\theta\in\Theta} T(\theta;\mathbf{x}).
\end{aligned}\right.
\end{equation}
In this paper, target detection is framed as a binary hypothesis test ($\mathcal{H}_0$: User A absent versus $\mathcal{H}_1$: User A present). The BS sensing receiver declares a detection if $T_{\max}(\mathbf{x}) > \tau$, where the sensing threshold $\tau$ is rigorously calibrated under $\mathcal{H}_0$ to satisfy a target probability of false alarm ($P_{\mathrm{FA}}$).

For performance evaluation, we focus on metrics that jointly capture detection reliability and localization precision:
\begin{align}
P_{\mathrm{D}} &\triangleq \Pr\big(T_{\max}(\mathbf{x})>\tau \mid \mathcal{H}_1\big), \label{eq:Pd}\\
P_{\mathrm{det} \wedge \mathrm{correct}} &\triangleq \Pr\big(T_{\max}(\mathbf{x})>\tau,\;\hat{\theta}=\theta_A \mid \mathcal{H}_1\big), \label{eq:Pdetcorr}\\
P_{\mathrm{correct} \mid \mathrm{detected}}
&\triangleq \frac{P_{\mathrm{det} \wedge \mathrm{correct}}}{P_{\mathrm{D}}}. \label{eq:Pcorr_given_det}
\end{align}
Additionally, we define the conditional localization root-mean-square error (RMSE) among valid detections:
\begin{equation}\label{eq:rmse_det}
\mathrm{RMSE}_{\mathrm{det}} \triangleq \sqrt{\mathbb{E}\big[ d_{\mathrm{circ}}(\hat{\theta},\theta_A)^2 \mid T_{\max}(\mathbf{x})>\tau,\mathcal{H}_1\big]},
\end{equation}
where $d_{\mathrm{circ}}(\hat{\theta},\theta) \triangleq \mathrm{wrap}\big(\hat{\theta}-\theta; N_\theta\big)$ computes the shortest circular distance on the modular grid.

\vspace{-2mm}
\subsection{Problem Formulation: Sensing-Centric Use Cases}\label{subsec:problems}
{\em Use Case I: Single-user stealth (clutter-likeness)}---In the absence of a target ($\mathcal{H}_0$), User B aims to suppress its own detectability. A principled mathematical notion of ``clutter-likeness'' is the statistical separation between the $\mathcal{H}_0$ distribution ($\mathbf{x}\sim\mathcal{CN}(\mathbf{0},\mathbf{R}_v)$) and the $\mathcal{H}_1$ distribution driven by User B's return ($\mathbf{x}\sim\mathcal{CN}(\mathbf{\mu}(\mathbf{q}),\mathbf{R}_v)$, where $\mathbf{\mu}(\mathbf{q})=\alpha_B(\mathbf{q})\mathbf{s}_{\theta_B}$). 

The Kullback-Leibler (KL) divergence between these distributions is precisely the whitened quadratic form:
\begin{equation}\label{eq:kl_mean_shift}
D_{\mathrm{KL}}\!\left(\mathcal{CN}(\mathbf{\mu}(\mathbf{q}),\mathbf{R}_v)\,\big\|\,\mathcal{CN}(\mathbf{0},\mathbf{R}_v)\right)
= |\alpha_B(\mathbf{q})|^2\,\mathbf{s}_{\theta_B}^{\mathsf{H}}\mathbf{R}_v^{-1}\mathbf{s}_{\theta_B}.
\end{equation}
Since $\mathbf{s}_{\theta_B}^{\mathsf{H}}\mathbf{R}_v^{-1}\mathbf{s}_{\theta_B}$ is structurally fixed, reducing detectability is equivalent to minimizing the effective coupling magnitude. The discrete port-selection problem is thus formulated as
\begin{equation}\label{eq:prob_stealth}
\begin{aligned}
\textbf{(P1) }\;\;\min_{\mathbf{q}\in\{0,1\}^K}\quad & \big|\mathbf{g}_B^{\mathsf{H}}\boldsymbol{\rho}(\mathbf{q})\big|^2\\
\text{s.t.}\quad & \mathbf{1}^{\mathsf{T}}\mathbf{q} = M_{\mathrm{active}}.
\end{aligned}
\end{equation}

{\em Use Case II: Cooperative interference shaping}---When localizing User A, User B should shape its spatial footprint to mitigate interference into A's localization neighborhood. We define a circular guard set encompassing $G$ bins around the target, $\mathcal{G}(\theta_A;G) \triangleq \big\{(\theta_A+\delta)\bmod N_\theta:\; \delta\in\{-G,\dots,G\}\big\}$.

User B's aggregate interference on this guard set relies on the whitened matched-filter overlap kernel, $\gamma(\theta,\theta_B) \triangleq \frac{|\mathbf{s}_\theta^{\mathsf{H}}\mathbf{R}_v^{-1}\mathbf{s}_{\theta_B}|^2}{\mathbf{s}_\theta^{\mathsf{H}}\mathbf{R}_v^{-1}\mathbf{s}_\theta}$. The objective is to suppress the energy of User B explicitly weighted by this overlap footprint:
\begin{equation}\label{eq:prob_coop}
\begin{aligned}
\textbf{(P2) }\;\;\min_{\mathbf{q}\in\{0,1\}^K}\quad & |\alpha_B(\mathbf{q})|^2 \sum_{\theta\in\mathcal{G}(\theta_A;G)} \gamma(\theta,\theta_B)\\
\text{s.t.}\quad & \mathbf{1}^{\mathsf{T}}\mathbf{q} = M_{\mathrm{active}}.
\end{aligned}
\end{equation}

{\em Combinatorial bottleneck and continuous relaxation}---Both \textbf{(P1)} and \textbf{(P2)} are non-convex mixed-integer non-linear programming (MINLP) problems. For FAS with $K$ ports activating a subset of size $M_{\mathrm{active}}$, the feasible discrete search space scales according to the binomial coefficient, $\binom{K}{M_{\mathrm{active}}}$. For massive FAS arrays (e.g., $K=200, M_{\mathrm{active}}=20$), this search space is astronomically large ($>10^{27}$ combinations). Furthermore, since the true channel $\mathbf{g}_B$ is only partially observable via $\mathbf{o}_B$, analytical solutions are impossible. Consequently, to solve these NP-hard problems in real time, Section \ref{sec:diffusion_solution} introduces a novel generative diffusion policy capable of sampling highly optimized masks $\mathbf{q}$ conditioned on the partial observations.

To facilitate real-time, AI-driven solutions, we map these discrete formulations into a continuous logit space, allowing for differentiable energy guidance. Also, we use the relaxed mask vector $\tilde{\mathbf{q}}\in[0,1]^K$ augmented with analytical penalty functions, \eqref{eq:E_card} and \eqref{eq:E_bin} (to be discussed later), to enforce cardinality and near-binary constraints during generation.

\vspace{-2mm}
\subsection{Diffusion-FAS Policy Interface}
The physical model above remains valid regardless of the algorithm used to compute the discrete mask $\mathbf{q}$. In the sequel, User B circumvents the combinatorial bottleneck by using a generative parameterization $\pi_\varphi$, mapping its side information $\mathbf{c}$ or context (including partial observation $\mathbf{o}_B$ and physical limits) into a distribution over valid masks:
\begin{equation}\label{eq:policy_def}
\mathbf{q} \sim \pi_\varphi(\mathbf{q} \mid \mathbf{c}).
\end{equation}
Specifically, we instantiate $\pi_\varphi$ through a conditional denoising diffusion model. This architecture enables the generation of highly diverse candidate masks that are dynamically steered toward the solutions of \textbf{(P1)} and \textbf{(P2)} through the continuous differentiable relaxations \eqref{eq:E_card} and \eqref{eq:E_bin}.

\vspace{-2mm}
\section{Energy-Guided Diffusion-FAS Policy}\label{sec:diffusion_solution}
To overcome the NP-hard combinatorial bottleneck formalized in Section~\ref{sec:system_model}, we propose \emph{Diffusion-FAS}, a conditional generative policy that enables a FAS-equipped object (User~B) to synthesize highly optimized multi-port state-coding masks. Depending on the scenario parameters, this policy seamlessly adapts to either (i) render the user's footprint statistically indistinguishable from background clutter (Use Case I: Stealth), or (ii) actively carve spatial nulls in the guard bins of a target-of-interest (Use Case II: Cooperative Shaping). 

Our architecture leverages a conditional DDPM. Instead of operating directly in the intractable discrete mask space, the DDPM diffuses over a continuous \emph{logit} space. During the reverse generative process, we employ a novel differentiable relaxation, allowing a customized sensing-energy function to guide the trajectory toward the optimal subset of $M_{\mathrm{active}}$ ports.

\vspace{-2mm}
\subsection{Continuous Logit Parameterization and Relaxation}\label{subsec:logits_to_mask}
We represent the discrete port state configuration as a binary mask $\mathbf{q} \in \{0,1\}^K$ subject to the cardinality constraint $\|\mathbf{q}\|_0 = M_{\mathrm{active}}$. To facilitate continuous gradient-based guidance, the diffusion model generates an unconstrained logit vector:
\begin{equation}
\mathbf{z} \in \mathbb{R}^K,
\end{equation}
which is mapped into either a \emph{hard mask} (for deployment and evaluation) or a \emph{soft mask} (for differentiable energy guidance).

{\em Hard top-$M_{\mathrm{active}}$ projection}---Given a generated logit vector $\mathbf{z}$, we define the index set of its $M_{\mathrm{active}}$ largest elements as $\mathcal{I}_{M_{\mathrm{active}}}(\mathbf{z}) \triangleq \operatorname{TopMIdx}(\mathbf{z})$. The deployment-ready hard mask is defined via the indicator function:
\begin{equation}\label{eq:hard_topm}
q_{k}(\mathbf{z}) \triangleq \mathbbm{1}\big\{k \in \mathcal{I}_{M_{\mathrm{active}}}(\mathbf{z})\big\},~k=1,\ldots,K.
\end{equation}

{\em Differentiable soft top-$M_{\mathrm{active}}$ relaxation}---As \eqref{eq:hard_topm} has zero gradients almost everywhere, we introduce a differentiable surrogate for the reverse diffusion process. Let $z_{(M_{\mathrm{active}})}$ be the $M_{\mathrm{active}}$-th largest scalar value in $\mathbf{z}$. We construct a continuous soft mask $\tilde{\mathbf{q}}(\mathbf{z}) \in [0,1]^K$ by applying a sharp sigmoid threshold centered at $z_{(M_{\mathrm{active}})}$:
\begin{equation}\label{eq:soft_topm}
\tilde{q}_{k}(\mathbf{z}) \triangleq \sigma\!\left(\frac{z_k - z_{(M_{\mathrm{active}})}}{\tau_q}\right),~\mbox{where }\sigma(u) = \frac{1}{1+e^{-u}},
\end{equation}
in which $\tau_q > 0$ controls the sharpness of the continuous approximation. This formulation pushes the top $M_{\mathrm{active}}$ logits toward $1$ and the remainder toward $0$, preserving the analytical gradients $\nabla_{\mathbf{z}} \tilde{\mathbf{q}}$ required by the diffusion solver.

{\em Continuous state-coding and effective coupling}---For any soft mask $\tilde{\mathbf{q}}$, the differentiable reflection vector is formed by continuously interpolating between the baseline and active impedance states $\boldsymbol{\rho}(\tilde{\mathbf{q}})$, see (\ref{eq:port-reflect-vector}). The continuous effective scalar coupling used for energy guidance is then given by \eqref{eq:effective_alphaB}.

\vspace{-2mm}
\subsection{Conditional Context Encoding}\label{subsec:context}
To ensure that the generative policy adapts dynamically to the instantaneous CSI and hardware constraints, our DDPM is conditioned on a context vector $\mathbf{c} \in \mathbb{R}^{d_c}$ which fuses the available partial observations and the system parameters:
\begin{equation}
\mathbf{c} = \big[\mathbf{o}_B^{\mathsf{T}}, \, \boldsymbol{\psi}^{\mathsf{T}}\big]^{\mathsf{T}},
\end{equation}
where $\mathbf{o}_B$ encodes the masked port-domain channel observed by User~B, see (\ref{eq:policy_obs}), and $\boldsymbol{\psi}$ embeds side information, such as the normalized physical constraints $(M_{\mathrm{active}}/K, M_{\mathrm{obs}}/K)$, the aperture size $W$, the background disturbance levels $(\sigma_c^2, \sigma_n^2)$, and (for Use Case II) the spatial locations of the targets.

\vspace{-2mm}
\subsection{Differentiable Energy Guidance Functions}\label{subsec:energy_guidance}
To steer the generative process toward the optimal masks for problems \textbf{(P1)} and \textbf{(P2)}, we construct a unified, differentiable energy functional $E(\tilde{\mathbf{q}})$. Using the whitened overlap kernel $\gamma(\theta, \theta_B)$ defined in Section~\ref{subsec:problems}, we define the following analytical energy sub-components:
\begin{align}
E_{\mathrm{int}}(\tilde{\mathbf{q}})
&\triangleq
|\alpha_B(\tilde{\mathbf{q}})|^2 \sum_{\theta\in\mathcal{G}(\theta_A;G)} \gamma(\theta, \theta_B), \label{eq:E_int} \\
E_{\mathrm{hide}}(\tilde{\mathbf{q}})
&\triangleq
|\alpha_B(\tilde{\mathbf{q}})|^2 \, \mathbf{s}_{\theta_B}^{\mathsf{H}}\mathbf{R}_v^{-1}\mathbf{s}_{\theta_B}, \label{eq:E_hide} \\
E_{\mathrm{card}}(\tilde{\mathbf{q}})
&\triangleq
\big(\mathbf{1}^\mathsf{T}\tilde{\mathbf{q}} - M_{\mathrm{active}}\big)^2, \label{eq:E_card} \\
E_{\mathrm{bin}}(\tilde{\mathbf{q}})
&\triangleq
\frac{1}{K}\sum_{k=1}^{K}\tilde{q}_{k}(1-\tilde{q}_{k}). \label{eq:E_bin}
\end{align}
Equations \eqref{eq:E_int} and \eqref{eq:E_hide} are the direct continuous relaxations of the ISAC objectives for Use Case II and Use Case I, respectively. Then \eqref{eq:E_card} and \eqref{eq:E_bin} act as physical-layer regularizers enforcing the cardinality budget and binary saturation. The total differentiable guidance energy is constructed as
\begin{multline}\label{eq:guidance_energy}
E(\tilde{\mathbf{q}}) \triangleq
\lambda_{\mathrm{int}}E_{\mathrm{int}}(\tilde{\mathbf{q}})
+
\lambda_{\mathrm{hide}}E_{\mathrm{hide}}(\tilde{\mathbf{q}})
+\\
\lambda_{\mathrm{card}}E_{\mathrm{card}}(\tilde{\mathbf{q}})
+
\lambda_{\mathrm{bin}}E_{\mathrm{bin}}(\tilde{\mathbf{q}}).
\end{multline}
By tuning the scalar weights, the identical diffusion architecture natively transitions between \textbf{Stealth} ($\lambda_{\mathrm{int}}=0, \lambda_{\mathrm{hide}}>0$) and \textbf{Cooperative Shaping} ($\lambda_{\mathrm{int}}>0, \lambda_{\mathrm{hide}} \ge 0$).

\vspace{-2mm}
\subsection{Forward Diffusion Process}\label{subsec:ddpm_forward}
We adopt a variance-preserving forward process. Let $\mathbf{z}_0 \in \mathbb{R}^K$ be an optimal ``clean'' logit vector for a given context $\mathbf{c}$. The forward diffusion process injects Gaussian noise over $T$ timesteps to produce a sequence of latent states $\{\mathbf{z}_t\}_{t=1}^{T}$:
\begin{equation}
q(\mathbf{z}_t \mid \mathbf{z}_{t-1})=\mathcal{N}\!\big(\sqrt{\alpha_t}\mathbf{z}_{t-1},(1-\alpha_t)\mathbf{I}\big),~t=1,\ldots,T,
\end{equation}
where $\alpha_t \triangleq 1-\beta_t$, and $\{\beta_t \in (0,1)\}_{t=1}^T$ is a predefined variance schedule. Defining $\bar{\alpha}_t \triangleq \prod_{i=1}^t \alpha_i$, the closed-form marginal distribution at any arbitrary timestep $t$ is
\begin{equation}\label{eq:ddpm_marginal}
q(\mathbf{z}_t \mid \mathbf{z}_0)=\mathcal{N}\!\big(\sqrt{\bar{\alpha}_t}\mathbf{z}_0,(1-\bar{\alpha}_t)\mathbf{I}\big).
\end{equation}
Equivalently, a noisy sample can be drawn directly as $\mathbf{z}_t = \sqrt{\bar{\alpha}_t}\mathbf{z}_0 + \sqrt{1-\bar{\alpha}_t}\,\boldsymbol{\epsilon}$, where $\boldsymbol{\epsilon}\sim\mathcal{N}(\mathbf{0},\mathbf{I})$. For large $T$, $\mathbf{z}_T$ approximates an isotropic Gaussian distribution, $\mathcal{N}(\mathbf{0},\mathbf{I})$.

\vspace{-2mm}
\subsection{Conditional Denoiser and Expert-Imitation Training}\label{subsec:ddpm_training}

The reverse generative process requires a neural network for predicting and removing the injected noise. We implement a conditional denoiser $\boldsymbol{\epsilon}_\phi(\mathbf{z}_t, t, \mathbf{c})$ parameterized by weights $\phi$. This network utilizes a multi-layer perceptron (MLP) backbone coupled with sinusoidal temporal embeddings for $t$. 

To bypass the necessity of online reinforcement learning, we train $\phi$ offline via \emph{expert-imitation}. We first generate a dataset $\mathcal{D}=\{(\mathbf{c}^{(i)}, \mathbf{z}_0^{(i)})\}_{i=1}^{N}$ using a heavily randomized heuristic oracle (detailed in Algorithm~\ref{alg:expert_data}) that exhaustively searches for high-quality logit vectors minimizing \eqref{eq:guidance_energy} given full CSI. 

The diffusion model is then trained to predict the noise $\boldsymbol{\epsilon}$ added to these expert logits using the DDPM objective
\begin{equation}\label{eq:ddpm_loss}
\mathcal{L}(\phi) =
\mathbb{E}_{\mathbf{z}_0,\mathbf{c},t,\boldsymbol{\epsilon}}
\left[
\left\|
\boldsymbol{\epsilon} -
\boldsymbol{\epsilon}_\phi\big(\sqrt{\bar{\alpha}_t}\mathbf{z}_0 + \sqrt{1-\bar{\alpha}_t}\boldsymbol{\epsilon},\, t,\, \mathbf{c}\big)
\right\|_2^2
\right].
\end{equation}

\begin{algorithm}[H]
\caption{\small Expert-Imitation Dataset Generation}\label{alg:expert_data}
{\footnotesize
\begin{algorithmic}[1]
\Require dataset size $N$; port count $K$; objective weights $(\lambda_{\mathrm{int}},\lambda_{\mathrm{hide}},\lambda_{\mathrm{card}},\lambda_{\mathrm{bin}})$; uniform ranges for constraints $[M_{\mathrm{active},\min},M_{\mathrm{active},\max}]$, $[M_{\mathrm{obs},\min},M_{\mathrm{obs},\max}]$.
\For{$i=1$ to $N$}
\State Sample limits $M_{\mathrm{active}} \sim \mathcal{U}[M_{\mathrm{active},\min}, M_{\mathrm{active},\max}]$ and $M_{\mathrm{obs}} \sim \mathcal{U}[M_{\mathrm{obs},\min}, M_{\mathrm{obs},\max}]$.
\State Sample physical scenario $(\mathbf{g}_B, \theta_A, \theta_B)$ and masking vector $\mathbf{m}$.
\State Construct context $\mathbf{c}^{(i)} = \big[\mathbf{o}_B^{\mathsf{T}}, \boldsymbol{\psi}^{\mathsf{T}}\big]^{\mathsf{T}}$.
\State Initialize $N_{\mathrm{cand}}$ continuous logit candidates $\{\mathbf{z}^{(j)}\}_{j=1}^{N_{\mathrm{cand}}}$ using random perturbations.
\State Project to hard masks $\mathbf{q}^{(j)}$ via \eqref{eq:hard_topm} and evaluate $E^{(j)}$ via \eqref{eq:guidance_energy}.
\State Identify global minimum: $j^\star = \arg\min_j E^{(j)}$.
\State Record target label: $\mathbf{z}_0^{(i)} = \mathbf{z}^{(j^\star)}$.
\EndFor
\Return $\mathcal{D}=\{(\mathbf{c}^{(i)},\mathbf{z}_0^{(i)})\}_{i=1}^{N}$.
\end{algorithmic}}
\end{algorithm}
{\em Robustness to hardware constraints}---A critical advantage of this framework is generalization. By randomizing the cardinality budget $M_{\mathrm{active}}$ and the observation budget $M_{\mathrm{obs}}$ during dataset generation and injecting them into the context vector $\mathbf{c}$, the neural network learns a universal policy. This eliminates the need to retrain the diffusion model for every sweep of hardware specifications in Section~\ref{sec:numerical_results}.

\vspace{-2mm}
\subsection{Energy-Guided Reverse Sampling}\label{subsec:ddpm_reverse}
At inference time, User B lacks full CSI and must generate an optimal mask relying strictly on the learned prior $p_\phi(\mathbf{z}_0 \mid \mathbf{c})$ and the partial context $\mathbf{c}$. We initialize the reverse process with pure noise $\mathbf{z}_T \sim \mathcal{N}(\mathbf{0},\mathbf{I})$. The unguided reverse transition is subsequently modeled as
\begin{equation}
p_\phi(\mathbf{z}_{t-1} \mid \mathbf{z}_t, \mathbf{c})=\mathcal{N}\!\big(\boldsymbol{\mu}_\phi(\mathbf{z}_t,t,\mathbf{c}),\tilde{\beta}_t\mathbf{I}\big),
\end{equation}
where the predicted reverse mean relies on the denoiser:
\begin{equation}\label{eq:reverse_mean}
\boldsymbol{\mu}_\phi(\mathbf{z}_t,t,\mathbf{c})=\frac{1}{\sqrt{\alpha_t}}\left(\mathbf{z}_t - \frac{\beta_t}{\sqrt{1-\bar{\alpha}_t}}\boldsymbol{\epsilon}_\phi(\mathbf{z}_t,t,\mathbf{c})\right),
\end{equation}
and $\tilde{\beta}_t \triangleq \beta_t\frac{1-\bar{\alpha}_{t-1}}{1-\bar{\alpha}_t}$ is the posterior variance.

\begin{algorithm}[]
\caption{\small Energy-Guided Reverse Sampling for Diffusion-FAS}\label{alg:guided_sample}
{\footnotesize
\begin{algorithmic}[]
\Require $\mathbf{c}$; total timesteps $T$; noise schedule $\{\beta_t\}$; guidance scale $\kappa$; candidate batch size $N_{\mathrm{cand}}$.
\For{$i=1$ to $N_{\mathrm{cand}}$}
\State Initialize pure noise latent state: $\mathbf{z}_T^{(i)} \sim \mathcal{N}(\mathbf{0},\mathbf{I})$.
\For{$t=T$ \textbf{down to} $1$}
\State Predict noise $\boldsymbol{\epsilon}_\phi(\mathbf{z}_t^{(i)},t,\mathbf{c})$ and unguided mean $\boldsymbol{\mu}_\phi$ via \eqref{eq:reverse_mean}.
\State Predict clean logits $\hat{\mathbf{z}}_0$ via \eqref{eq:z0_hat} and soft mask $\tilde{\mathbf{q}}$ via \eqref{eq:soft_topm}.
\State Backpropagate sensing energy gradient $\mathbf{g}_t = \nabla_{\mathbf{z}_t} E(\tilde{\mathbf{q}})$.
\State Nudge mean: $\tilde{\boldsymbol{\mu}}_\phi = \boldsymbol{\mu}_\phi - \kappa \tilde{\beta}_t \mathbf{g}_t$.
\State Sample $\boldsymbol{\eta} \sim \mathcal{N}(\mathbf{0},\mathbf{I})$ (if $t>1$, else $\boldsymbol{\eta}=\mathbf{0}$).
\State Execute reverse step: $\mathbf{z}_{t-1}^{(i)} = \tilde{\boldsymbol{\mu}}_\phi + \sqrt{\tilde{\beta}_t}\boldsymbol{\eta}$.
\EndFor
\State Extract final hard mask $\mathbf{q}^{(i)}$ from terminal state $\mathbf{z}_0^{(i)}$ via \eqref{eq:hard_topm}.
\State Evaluate true deterministic energy $E^{(i)}$.
\EndFor
\State \textbf{Select optimal solution:} $i^\star = \arg\min_i E^{(i)}$.
\Return Optimal hardware mask $\mathbf{q}^\star = \mathbf{q}^{(i^\star)}$.
\end{algorithmic}}
\end{algorithm}

To guarantee that the sampled mask strictly adheres to the sensing constraints, we augment this reverse step using \emph{energy-guidance} conceptually analogous to Langevin dynamics and classifier guidance in image diffusion. At each timestep $t$, we obtain a differentiable proxy of the clean logits:
\begin{equation}\label{eq:z0_hat}
\hat{\mathbf{z}}_0(\mathbf{z}_t, t, \mathbf{c})
=
\frac{1}{\sqrt{\bar{\alpha}_t}}
\left(
\mathbf{z}_t - \sqrt{1-\bar{\alpha}_t}\,\boldsymbol{\epsilon}_\phi(\mathbf{z}_t,t,\mathbf{c})
\right).
\end{equation}
We map $\hat{\mathbf{z}}_0$ into a soft mask $\tilde{\mathbf{q}}(\hat{\mathbf{z}}_0)$ via \eqref{eq:soft_topm} and compute the ISAC energy $E(\tilde{\mathbf{q}})$. Utilizing automatic differentiation, we compute the analytical gradient with respect to the latent state:
\begin{equation}\label{eq:energy_grad}
\mathbf{g}_t \triangleq \nabla_{\mathbf{z}_t} E\big(\tilde{\mathbf{q}}(\hat{\mathbf{z}}_0(\mathbf{z}_t,t,\mathbf{c}))\big).
\end{equation}
The reverse mean is nudged in the negative gradient direction:
\begin{equation}\label{eq:guided_mean}
\tilde{\boldsymbol{\mu}}_\phi=\boldsymbol{\mu}_\phi(\mathbf{z}_t,t,\mathbf{c})-\kappa\,\tilde{\beta}_t\,\mathbf{g}_t,
\end{equation}
where $\kappa \ge 0$ denotes the guidance scale. The guided reverse transition finally executes as $\mathbf{z}_{t-1} = \tilde{\boldsymbol{\mu}}_\phi + \sqrt{\tilde{\beta}_t}\,\boldsymbol{\eta}$, with $\boldsymbol{\eta} \sim \mathcal{N}(\mathbf{0},\mathbf{I})$. As the diffusion process is stochastic, we generate $N_{\mathrm{cand}}$ parallel candidates and ultimately select the hard mask $\mathbf{q}^\star$ that minimizes the deterministic energy functional. The whole procedure is formalized in Algorithm~\ref{alg:guided_sample}.

\vspace{-2mm}
\subsection{Computational Complexity}\label{subsec:algorithms}
The denoiser $\boldsymbol{\epsilon}_\phi$ is realized as an MLP mapping $\mathbb{R}^{K+d_c} \to \mathbb{R}^{K}$. As such, a single forward pass scales as $\mathcal{O}(K H)$, where $H$ is the hidden layer dimensionality. The guided sampling mechanism requires an additional backward pass per timestep to evaluate the Jacobian of \eqref{eq:energy_grad}, computationally tracked efficiently via reverse-mode automatic differentiation. The total complexity scales as $\mathcal{O}(N_{\mathrm{cand}} T \text{Cost(MLP)})$. Since the energy guidance function evaluates only the structural overlap tensor $\mathbf{S}^\mathsf{H}\mathbf{R}_v^{-1}\mathbf{S}$ and avoids tracking the exact instantaneous disturbance sample $\mathbf{v}$, the algorithmic execution strictly prevents unintended information leakage while remaining computationally feasible for symbol-by-symbol operations.


\vspace{-3mm}
\section{Numerical Results}\label{sec:numerical_results}
In this section, we assess the performance of the proposed Diffusion-FAS framework across two ISAC paradigms: Case 1 (Generative Spatial Stealth), where a FAS-equipped object minimizes its sensing visibility by navigating into localized deep fades; and Case 2 (ISAC Target Isolation), which stresses the system's ability to isolate target reflections from spatially adjacent, uncoordinated interferers under varying degrees of environmental clutter and spatial proximity.

\vspace{-2mm}
\subsection{System Settings}
We consider that User B is equipped with FAS comprising $K = 200$ dynamically selectable antenna ports packed within a default normalized physical aperture of $W = 2\lambda\times2\lambda$. In Case 1, the system objective is to minimize the probability of detection ($P_D$) for a single user, while in Case 2, the system must detect and localize a target (User A) in the presence of an uncoordinated interferer (User B). To demonstrate the robustness of the generative prior across diverse physical environments, both finite and rich scattering environments are considered. During the inference phase, the DDPM context vector is conditioned on $M_{\mathrm{obs}}$ randomly sampled observation ports to reconstruct the manifold. The primary system settings and DDPM hyperparameters are summarized in Table \ref{tab:sim_params}.

\begin{table}[]
\centering
\caption{Simulation Parameters and DDPM Hyperparameters}
\label{tab:sim_params}
\renewcommand{\arraystretch}{1.2}
\resizebox{.8\columnwidth}{!}{
\begin{tabular}{l l}
\hline\hline
\textbf{Parameter} & \textbf{Value} \\ 
\hline
\multicolumn{2}{c}{\textbf{FAS and ISAC Settings}} \\
\hline
Total number of FAS ports ($K$) & $200$ \\ 
Normalized aperture size ($W$) & $2\lambda\times2\lambda$ (Default) \\ 
Active ports ($M_{\mathrm{active}}$) & $1 \sim 60$ (Case 1), $20$ (Case 2) \\ 
Observed context ports ($M_{\mathrm{obs}}$) & $30$ (Default) \\ 
Spatial grid offset ($\Delta$) & N/A (Case 1), $0 \sim 16$ bins (Case 2) \\ 
Clutter standard deviation ($\sigma_c$) & $1.0$ (Default) \\ 
Target False Alarm Rate ($P_{\mathrm{FA}}$) & $0.01, 0.1$ (Case 1), $10^{-3}$ (Case 2) \\
\hline
\multicolumn{2}{c}{\textbf{Diffusion Model Hyperparameters}} \\
\hline
Diffusion timesteps ($T$) & $1000$ \\
Noise schedule & Linear ($\beta_1 = 10^{-4}, \beta_T = 0.02$) \\
Neural network architecture & Conditional MLP \\
Hidden dimensions & $[512, 1024, 1024, 512]$ \\
Optimizer & Adam ($LR = 1 \times 10^{-4}$) \\
Training batch size & $256$ \\
\hline\hline
\end{tabular}}

\end{table}

\vspace{-2mm}
\subsection{Baseline Schemes}
To benchmark the proposed generative framework, we compare \textbf{Diffusion-FAS} against three distinct baseline schemes.
\begin{itemize}
\item \textbf{No User B (Theoretical Upper Bound)}---This scheme evaluates the FAS performance in a pristine, interference-free vacuum where User B does not exist. The spatial correlation is solely driven by the target, serving as the absolute physical upper limit for detection probability.
\item \textbf{Random FAS (Uninformed Selection)}---In this scheme, the receiver randomly selects $M_{\mathrm{active}}$ ports uniformly across the $K$ available physical locations. Comparing the proposed policy against Random FAS strictly quantifies the gain achieved by \textit{intelligent, physics-informed} spatial sampling versus relying on blind spatial diversity.
\item \textbf{No FAS (Hardware Lower Bound)}---This scheme restricts the receiver to a traditional, static multi-antenna array format. The antennas are permanently fixed at uniformly spaced intervals ($M_{\mathrm{active}}$ elements spanning $W$). This baseline demonstrates the fundamental necessity of FAS in overcoming highly localized interference.
\end{itemize}

\vspace{-2mm}
\subsection{Case 2: ISAC Target Isolation and Interference Nulling}

\begin{figure}[]
\centering
\subfigure[]{\includegraphics[width=0.8\columnwidth]{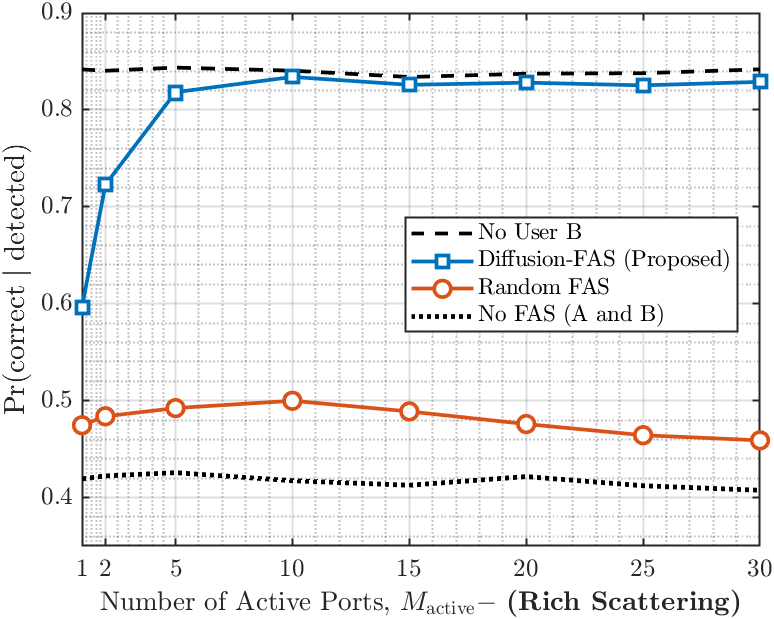}\label{fig:m_active_rich_cond}}
\subfigure[]{\includegraphics[width=0.8\columnwidth]{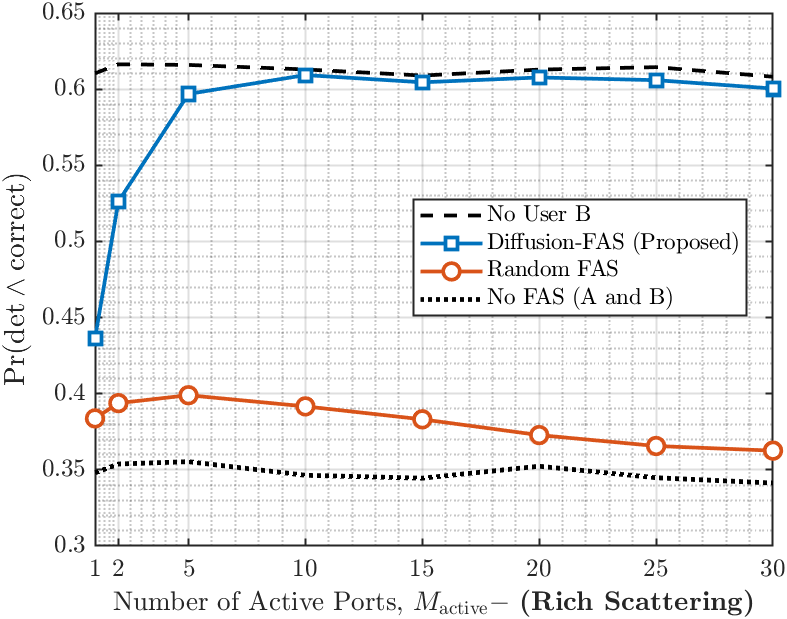}\label{fig:m_active_rich_joint}}
\caption{Impact of the number of active ports ($M_{\mathrm{active}}$) on localization and detection performance under rich isotropic scattering.}\label{fig:m_active_rich}
\vspace{-5mm}
\end{figure}

\begin{figure}[]
\centering
\subfigure[]{\includegraphics[width=0.8\columnwidth]{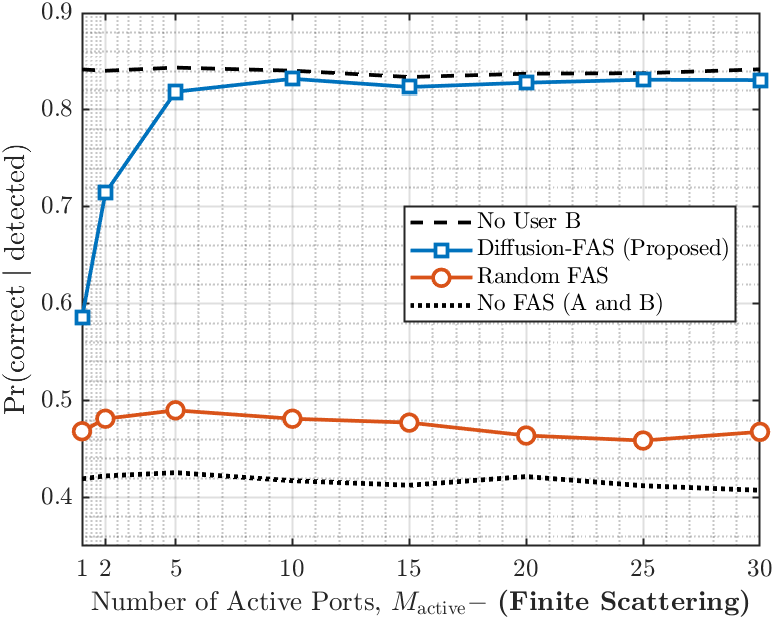}\label{fig:m_active_finite_cond}}
\subfigure[]{\includegraphics[width=0.8\columnwidth]{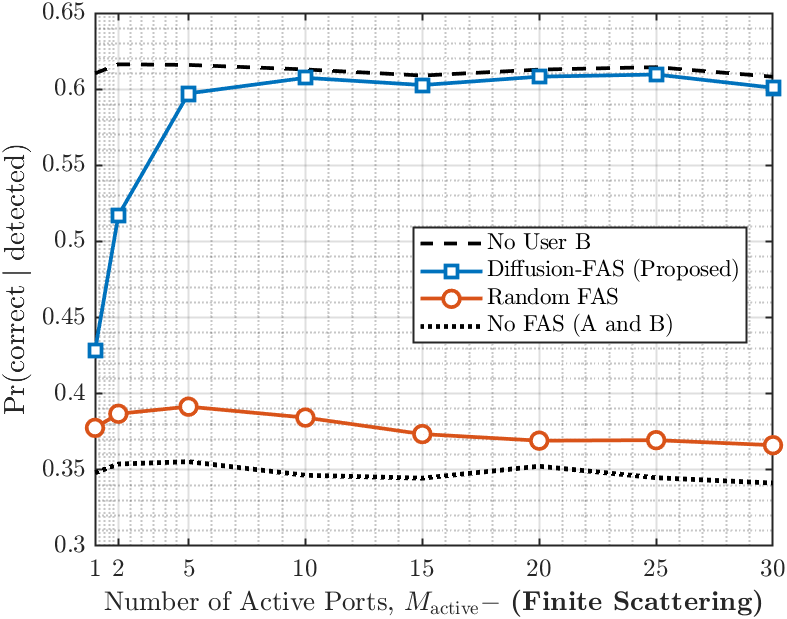}\label{fig:m_active_finite_joint}}
\caption{Impact of the number of active ports ($M_{\mathrm{active}}$) on localization and detection performance under finite scattering.}\label{fig:m_active_finite}
\vspace{-5mm}
\end{figure}

Figs.~\ref{fig:m_active_rich} and \ref{fig:m_active_finite} investigate the impact of the active port count ($M_{\mathrm{active}}$) on localization and joint detection accuracy for rich isotropic and finite scattering environments, respectively. As $M_{\mathrm{active}}$ expands from $1$ to $5$, the proposed Diffusion-FAS policy exhibits a sharp, non-linear performance gain, rapidly closing the performance gap with the theoretical interference-free upper bound (No User B). This steep ascent demonstrates the generative prior's ability to efficiently leverage emerging spatial DoFs to synthesize precise spatial nulls against the interferer. Crucially, the performance of Diffusion-FAS smoothly saturates at a highly compact active subset of $M_{\mathrm{active}} \approx 5$. In stark contrast, the Random FAS and static No FAS baselines remain stagnated at a lower performance floor, confirming that naive spatial sampling cannot overcome severe proximity interference, regardless of how many ports are activated.

\begin{figure}[]
\centering
\subfigure[]{\includegraphics[width=0.8\columnwidth]{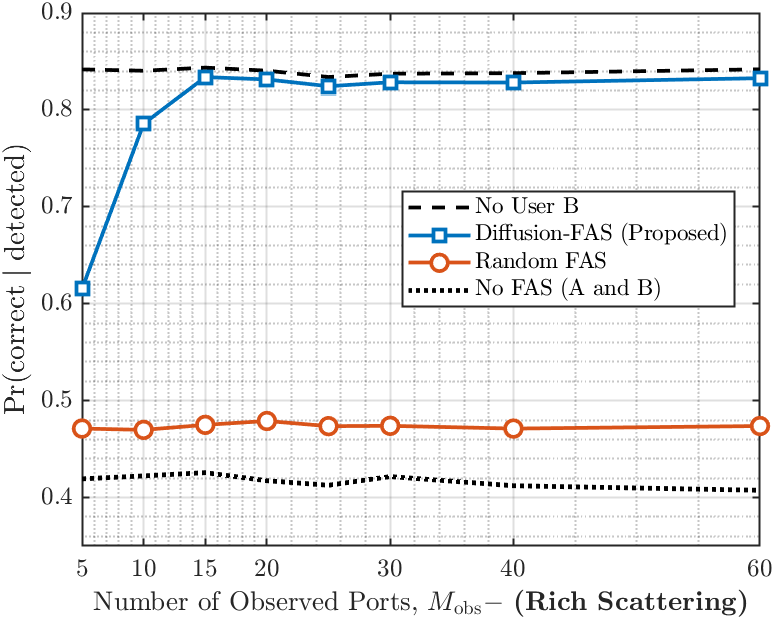}\label{fig:m_obs_rich_cond}}
\subfigure[]{\includegraphics[width=0.8\columnwidth]{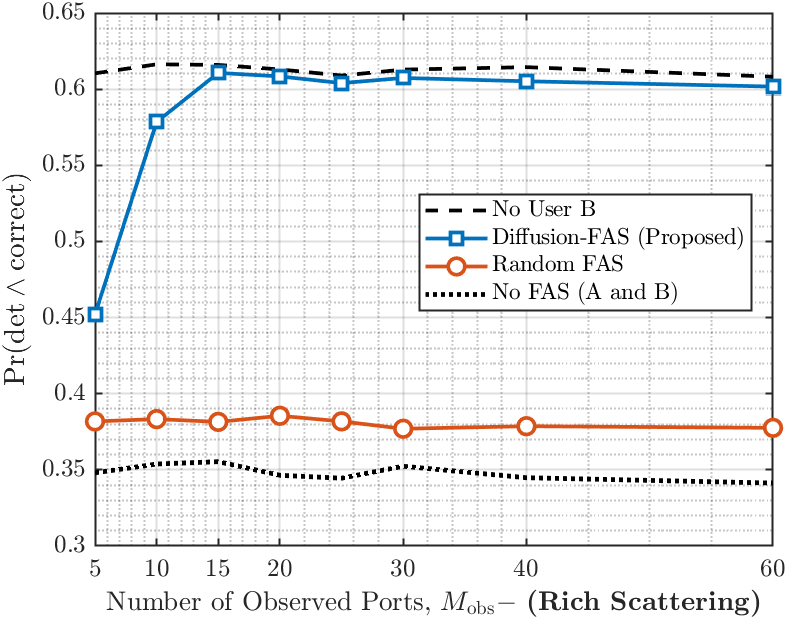}\label{fig:m_obs_rich_joint}}
\caption{Impact of the number of observed context ports ($M_{\mathrm{obs}}$) on localization and detection performance under rich isotropic scattering.}\label{fig:m_obs_rich}
\vspace{-5mm}
\end{figure}

\begin{figure}[]
\centering
\subfigure[]{\includegraphics[width=0.8\columnwidth]{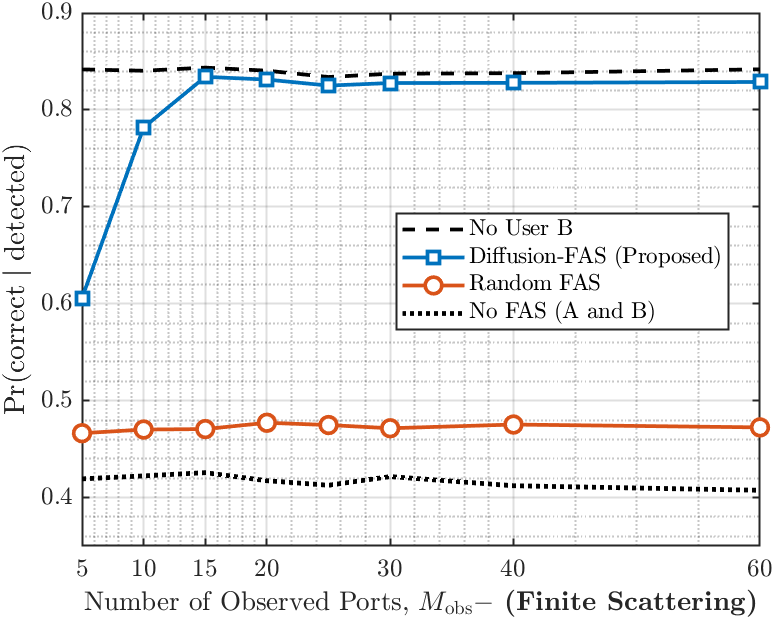}\label{fig:m_obs_finite_cond}}
\subfigure[]{\includegraphics[width=0.8\columnwidth]{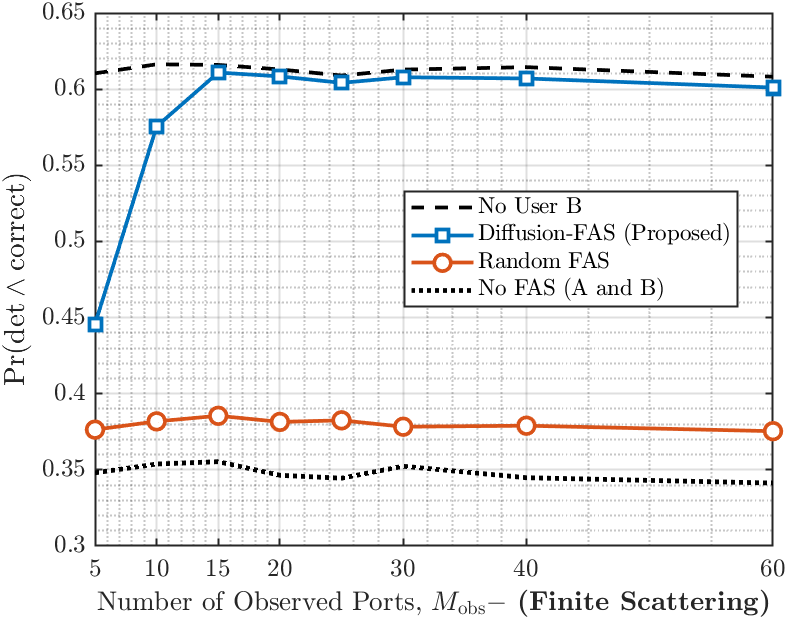}\label{fig:m_obs_finite_joint}}
\caption{Impact of the number of observed context ports ($M_{\mathrm{obs}}$) on localization and detection performance under finite scattering.}\label{fig:m_obs_finite}
\vspace{-5mm}
\end{figure}

Figs.~\ref{fig:m_obs_rich} and \ref{fig:m_obs_finite} evaluate the system's robustness to partial spatial observability ($M_{\mathrm{obs}}$) during the inference phase. Under extreme spatial sparsity ($M_{\mathrm{obs}} \le 10$), the context vector provides insufficient spatial anchor points to adequately resolve the high-frequency fluctuations of the multipath fading envelope. Consequently, the energy-guidance gradients occasionally steer the generative trajectory toward suboptimal local minima, resulting in degraded localization. However, as the observation window expands to $M_{\mathrm{obs}} \approx 15 \sim 20$, the Diffusion-FAS policy exhibits a sharp performance recovery, successfully locking onto the underlying spatial correlation structure and tightly converging with the theoretical No User B upper bound. This highlights a powerful capability of the generative approach. It can accurately reconstruct and exploit the full spatial interference field using less than $10\%$ of the available $K=200$ physical ports. Interestingly, a slight performance dip is observed at dense observability ($M_{\mathrm{obs}} = 60$), which is a well-documented artifact of over-conditioning in conditional generative models. In this regime, an overly dense context vector prematurely collapses the generative trajectory, overriding the exploratory stochastic noise needed to find the absolute global minimum. Predictably, the non-adaptive Random FAS and static No FAS baselines remain completely invariant to $M_{\mathrm{obs}}$ at a severely degraded performance floor.

\begin{figure}[]
\centering
\subfigure[]{\includegraphics[width=0.8\columnwidth]{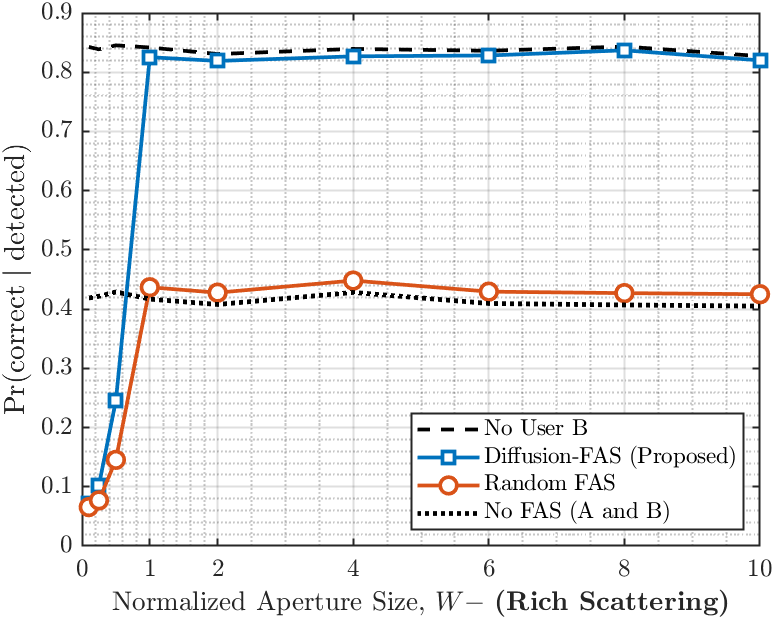}\label{fig:w_rich_cond}}
\subfigure[]{\includegraphics[width=0.8\columnwidth]{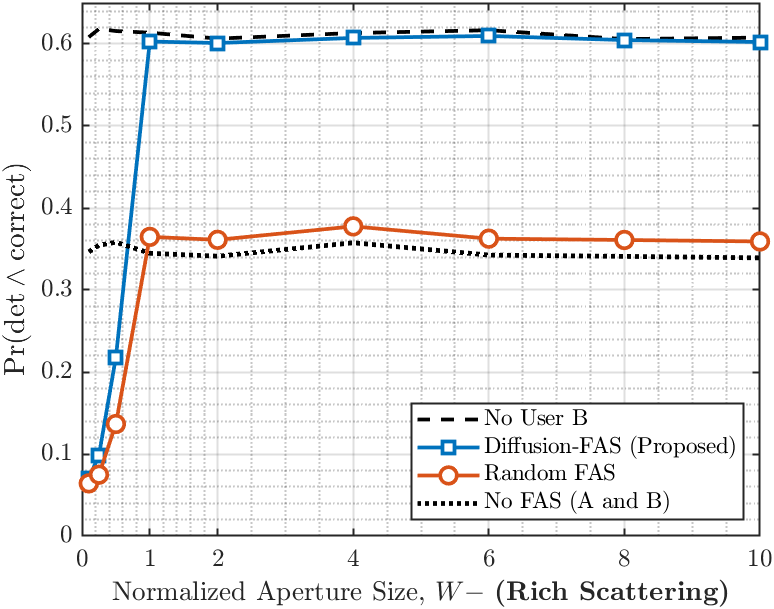}\label{fig:w_rich_joint}}
\caption{Impact of the normalized aperture size ($W$) on localization and detection performance under rich isotropic scattering.}\label{fig:w_rich}
\vspace{-5mm}
\end{figure}

\begin{figure}[]
\centering
\subfigure[]{\includegraphics[width=0.8\columnwidth]{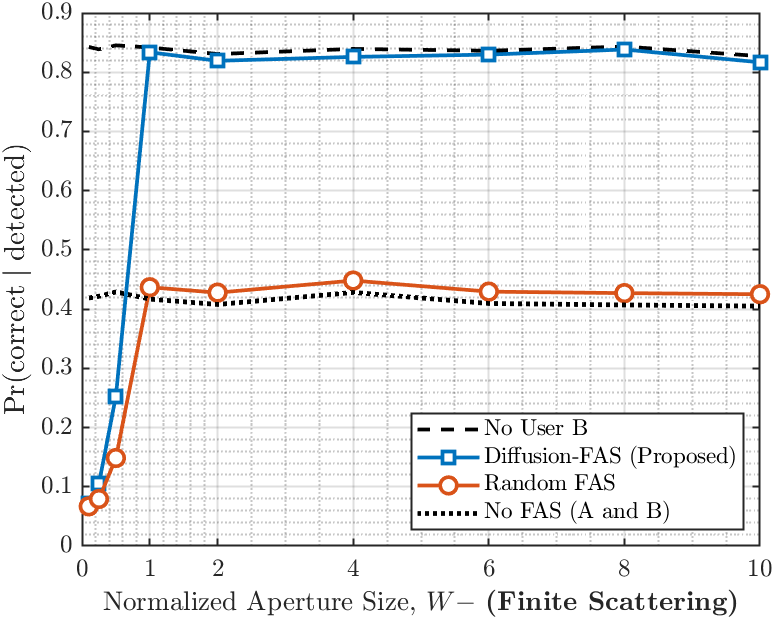}\label{fig:w_finite_cond}}
\subfigure[]{\includegraphics[width=0.8\columnwidth]{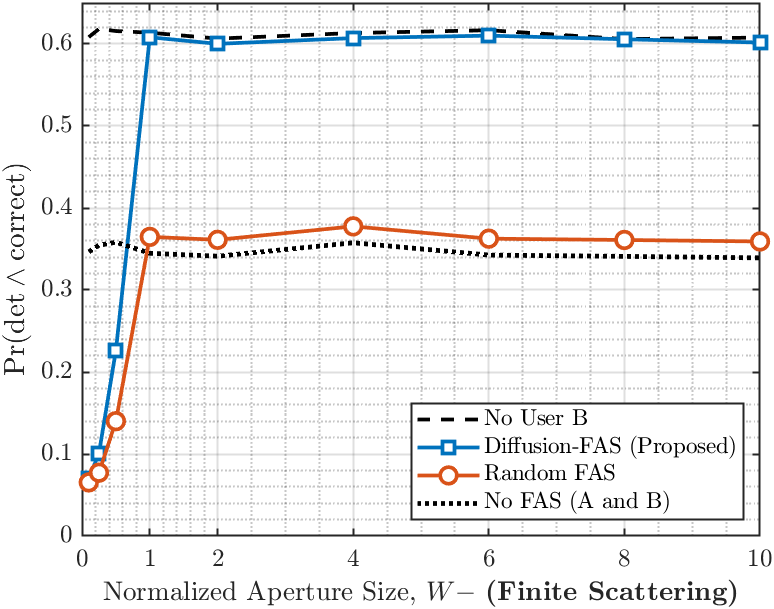}\label{fig:w_finite_joint}}
\caption{Impact of the normalized aperture size ($W$) on localization and detection performance under finite scattering.}\label{fig:w_finite}
\vspace{-5mm}
\end{figure}

\begin{figure}[]
\centering
\subfigure[]{\includegraphics[width=0.8\columnwidth]{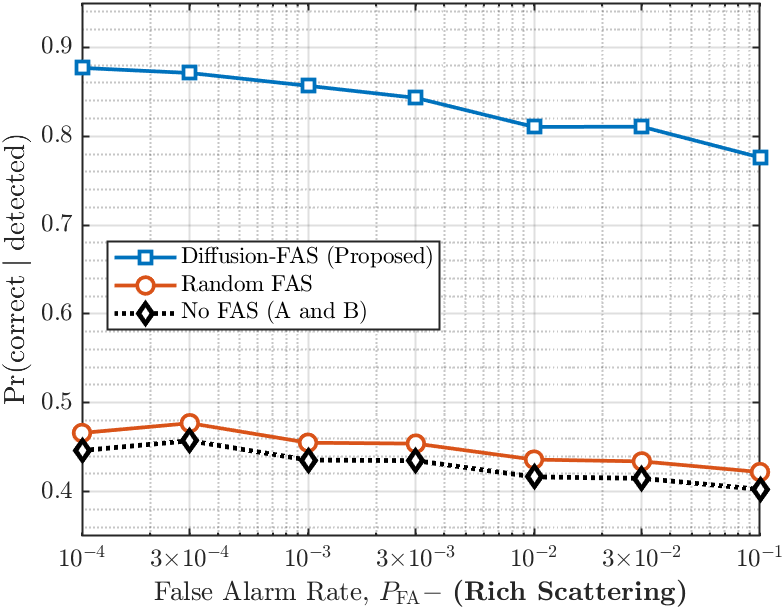}\label{fig:pfa_rich_cond}}
\subfigure[]{\includegraphics[width=0.8\columnwidth]{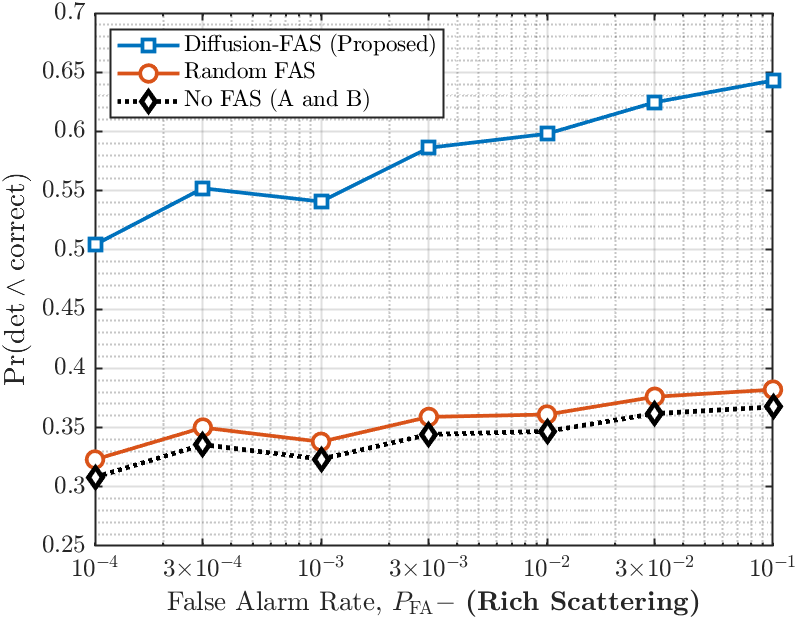}\label{fig:pfa_rich_joint}}
\caption{Impact of the target false alarm rate ($P_{\mathrm{FA}}$) on localization and detection performance under rich isotropic scattering.}\label{fig:pfa_rich}
\vspace{-5mm}
\end{figure}

Figs.~\ref{fig:w_rich} and \ref{fig:w_finite} study the system's localization and detection performance across varying normalized aperture sizes ($W$), focusing on the critical transition from the sub-wavelength regime to macroscopic arrays. At an extreme sub-wavelength footprint ($W \le 0.5\lambda$), the $K=200$ fluid ports are densely packed within a minuscule area. Consequently, the spatial correlation among the ports approaches unity. Denied any meaningful spatial diversity, both Diffusion-FAS and Random FAS experience a severe performance collapse, regressing toward the static No FAS lower bound. Nevertheless, as the aperture expands beyond $0.5\lambda$, the spatial manifold rapidly decorrelates. The generative prior seamlessly leverages these emerging spatial DoFs, causing the localization accuracy to rise steeply. The performance of the Diffusion-FAS policy smoothly saturates at $W \ge 2\lambda$, converging to the interference-free No User B upper bound. This confirms that cooperative FAS interference shaping does not require massive, unwieldy form factors. A highly compact $2\lambda$ footprint encapsulates sufficient spatial DoFs for AI to fully resolve the multipath envelope and synthesize optimal spatial nulls.

Fig.~\ref{fig:pfa_rich} studies the receiver operating characteristic (ROC) behavior of the ISAC system by evaluating performance across a sweeping target false alarm rate ($P_{\mathrm{FA}}$). This sweep illustrates the fundamental trade-off between detection sensitivity and spatial localization accuracy. As the CFAR threshold relaxes (i.e., $P_{\mathrm{FA}}$ increases from $10^{-4}$ to $10^{-1}$), the conditional correctness, $\Pr(\mathrm{correct} \mid \mathrm{detected})$, naturally degrades across all schemes. This occurs because a more permissive threshold admits a higher volume of noise and residual interference spikes, increasing the likelihood of false spatial peaks. However, the proposed Diffusion-FAS maintains a commanding performance gap over the baselines across the entire sweep, proving its superior interference-nulling capabilities even under strict threshold constraints. Conversely, the joint detection probability, $\Pr(\mathrm{det} \wedge \mathrm{correct})$, rises as the relaxed threshold increases the raw probability of detection ($P_{\mathrm{D}}$). Even in this metric, Diffusion-FAS scales significantly better than the static and random baselines. This confirms that the generative policy structurally improves the fundamental radar performance bounds, ensuring that when the system registers a detection, it is overwhelmingly more likely to be spatially accurate compared to traditional spatial sampling methods.

\begin{figure}[]
\centering
\subfigure[]{\includegraphics[width=0.8\columnwidth]{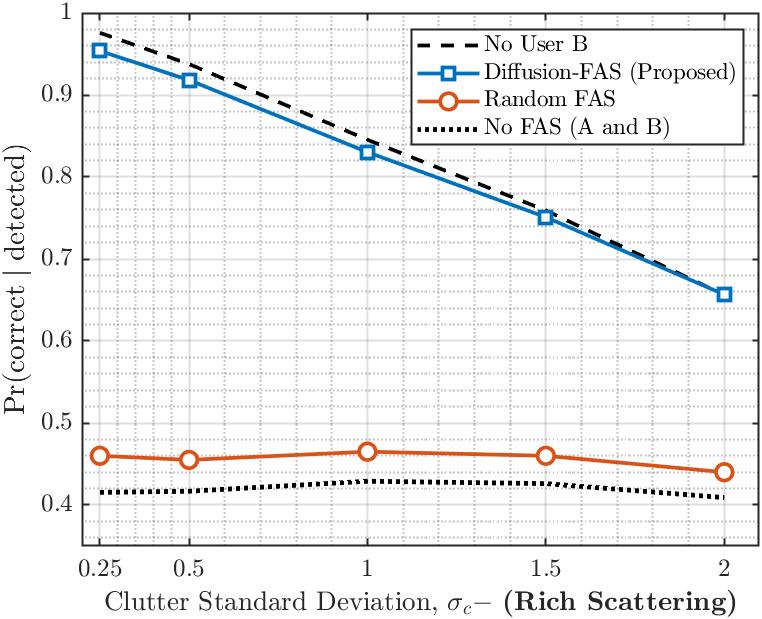}\label{fig:clutter_rich_cond}}
\subfigure[]{\includegraphics[width=0.8\columnwidth]{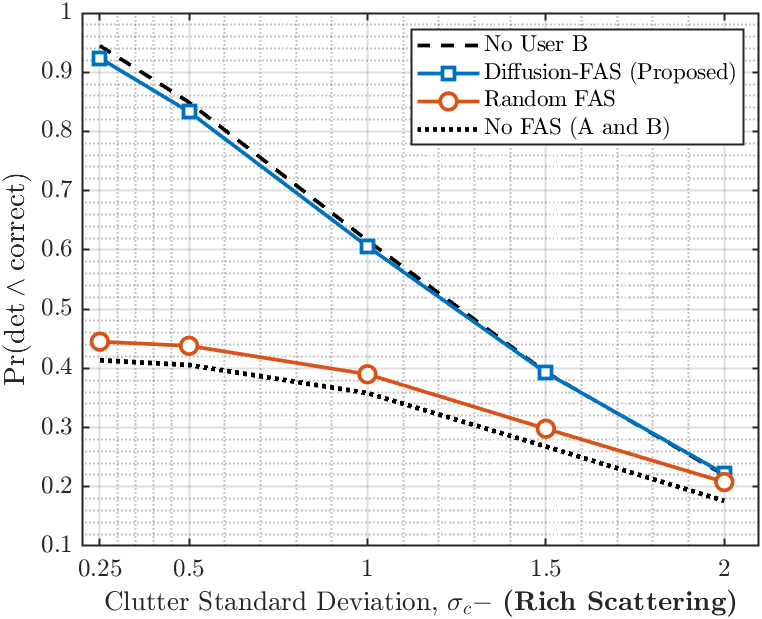}\label{fig:clutter_rich_joint}}
\caption{Impact of the clutter standard deviation ($\sigma_c$) on localization and detection performance under rich isotropic scattering.}\label{fig:clutter_rich}
\vspace{-5mm}
\end{figure}

\begin{figure}[]
\centering
\subfigure[]{\includegraphics[width=0.8\columnwidth]{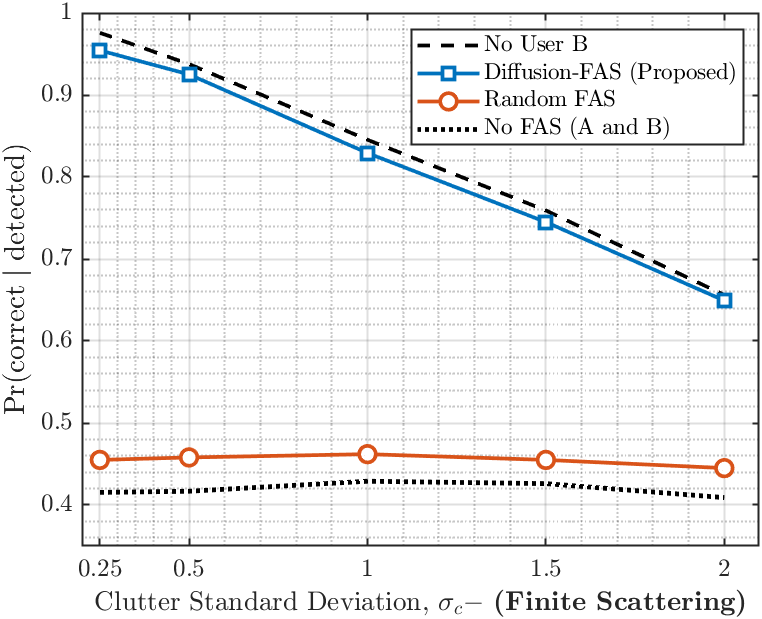}\label{fig:clutter_finite_cond}}
\subfigure[]{\includegraphics[width=0.8\columnwidth]{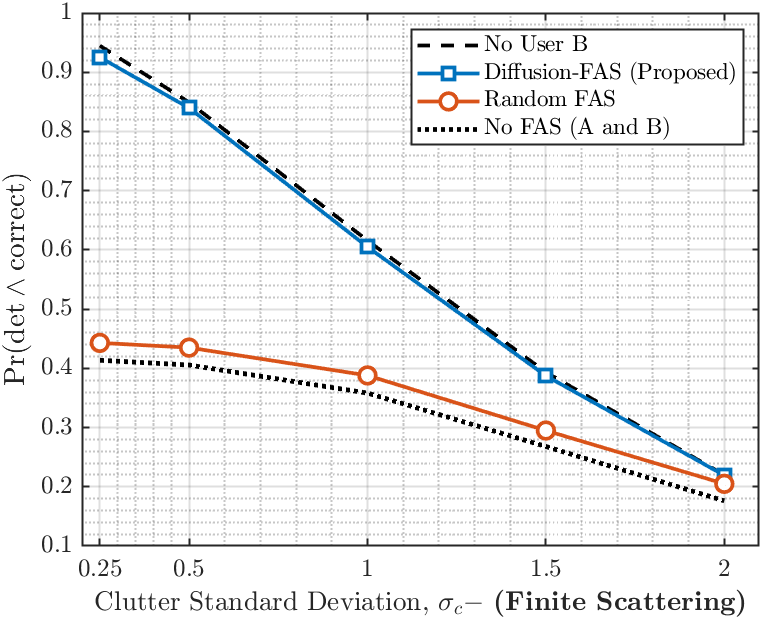}\label{fig:clutter_finite_joint}}
\caption{Impact of the clutter standard deviation ($\sigma_c$) on localization and detection performance under finite scattering.}\label{fig:clutter_finite}
\vspace{-5mm}
\end{figure}

In Figs.~\ref{fig:clutter_rich} and \ref{fig:clutter_finite}, we assess the system's robustness against varying degrees of clutter severity ($\sigma_c$). As the clutter standard deviation increases from $0.25$ to $2$, the localization and joint detection probabilities naturally experience a monotonic decline across all schemes. This is a fundamental physical limitation; extreme clutter severely masks target reflections, distorts the spatial envelope, and introduces dense false alarms, directly dragging down even the theoretical interference-free upper bound (No User B). However, despite this harsh environmental degradation, the proposed Diffusion-FAS policy exhibits remarkable resilience. It tightly tracks the No User B bound across the entire spectrum, effectively neutralizing the adjacent interferer regardless of the background clutter intensity. Conversely, the non-adaptive Random FAS and No FAS baselines are severely crippled from the outset. Because they lack the intelligence to form targeted spatial nulls, their performance is instantly dominated by the uncoordinated interference, relegating them to a heavily degraded performance floor even in benign, low-clutter environments ($\sigma_c = 0.25$). 

\begin{figure}[]
\centering
\subfigure[]{\includegraphics[width=0.8\columnwidth]{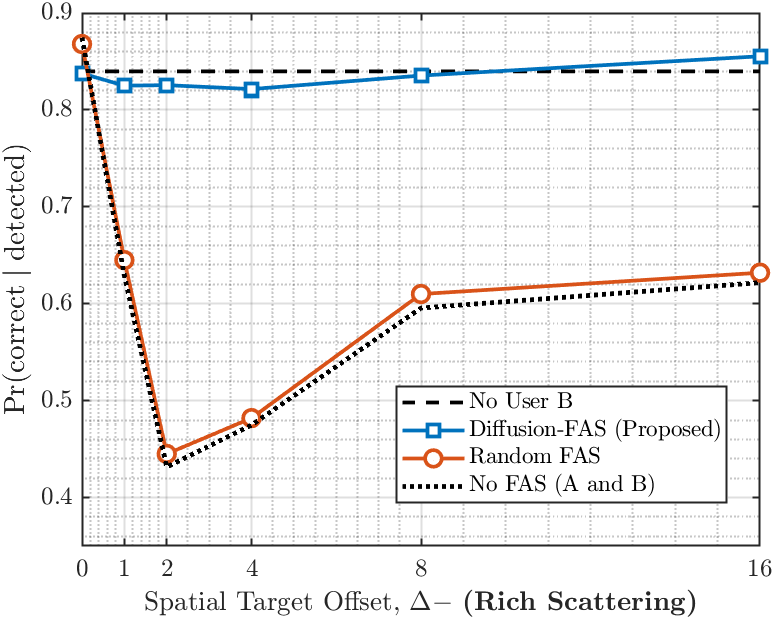}\label{fig:delta_rich_cond}}
\subfigure[]{\includegraphics[width=0.8\columnwidth]{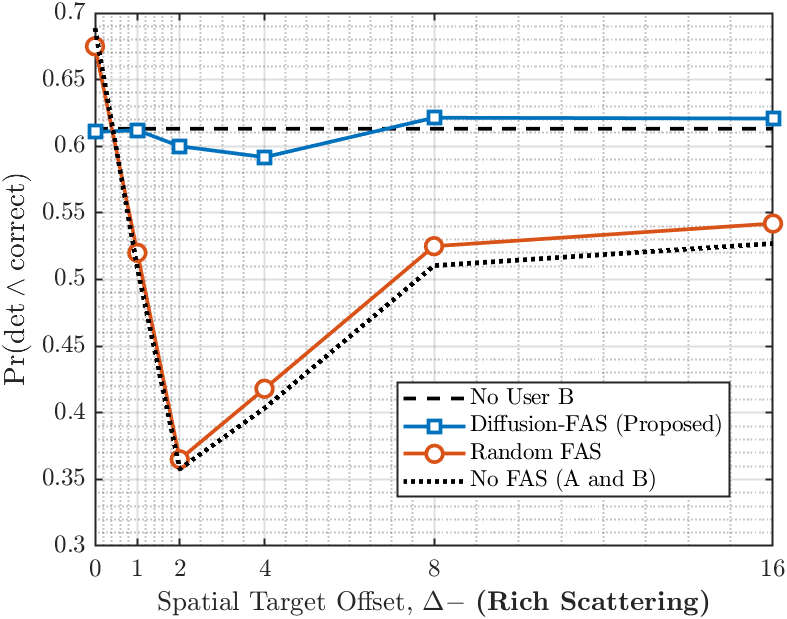}\label{fig:delta_rich_joint}}
\caption{Impact of the spatial target offset ($\Delta$) on localization and detection performance under rich isotropic scattering.}
\label{fig:delta_rich}
\vspace{-5mm}
\end{figure}

\begin{figure}[]
\centering
\subfigure[]{\includegraphics[width=0.8\columnwidth]{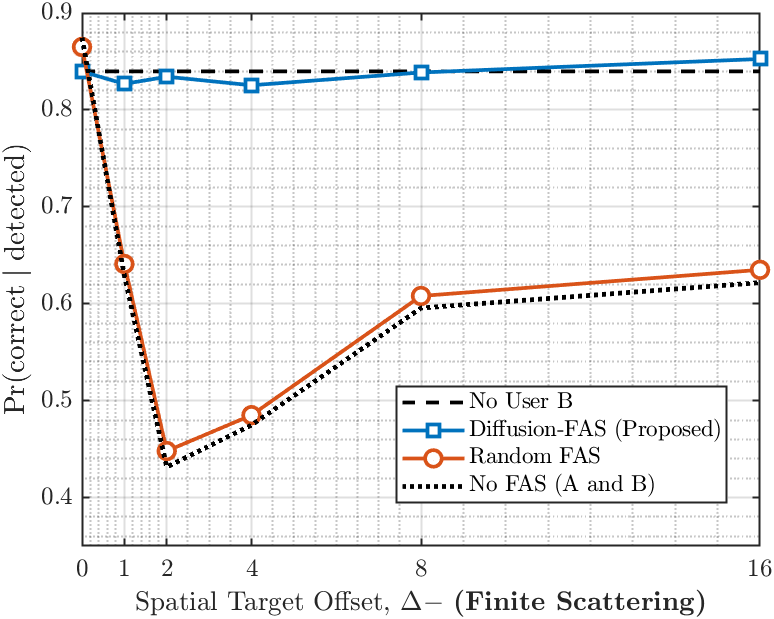}\label{fig:delta_finite_cond}}
\subfigure[]{\includegraphics[width=0.8\columnwidth]{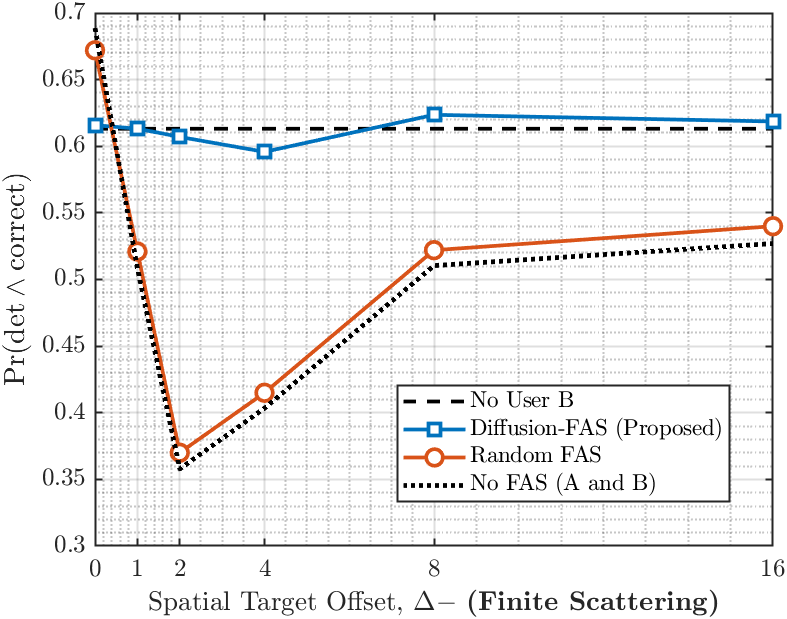}\label{fig:delta_finite_joint}}
\caption{Impact of the spatial target offset ($\Delta$) on localization and detection performance under finite scattering.}\label{fig:delta_finite}
\vspace{-5mm}
\end{figure}

Figs.~\ref{fig:delta_rich} and \ref{fig:delta_finite} examine the impact of interference geometry by varying the spatial offset ($\Delta$) between the target and the interferer. A unique physical phenomenon is observed at $\Delta = 0$, where the interferer and target occupy the exact same spatial grid point. In this singular case, the interferer effectively acts as a constructive beacon, dumping uncoordinated energy into the true target bin and causing the baseline schemes to exhibit misleadingly high localization accuracy. However, as the interferer shifts just a single bin away ($\Delta \ge 1$), the baseline Random FAS and static No FAS schemes experience a catastrophic performance collapse, with accuracy dropping below $20\%$. This occurs because their lack of spatial nulling capability allows the interferer's energy to completely mask the target's spatial signature. In contrast, the proposed Diffusion-FAS policy demonstrates remarkable robustness to proximity; it actively synthesizes a spatial null at the interferer's location, maintaining a stable performance profile that tracks the theoretical No User B upper bound. 

\vspace{-2mm}
\subsection{Case 1: Generative Spatial Stealth Performance}

\begin{figure}[]
\centering
\subfigure[]{\includegraphics[width=0.8\columnwidth]{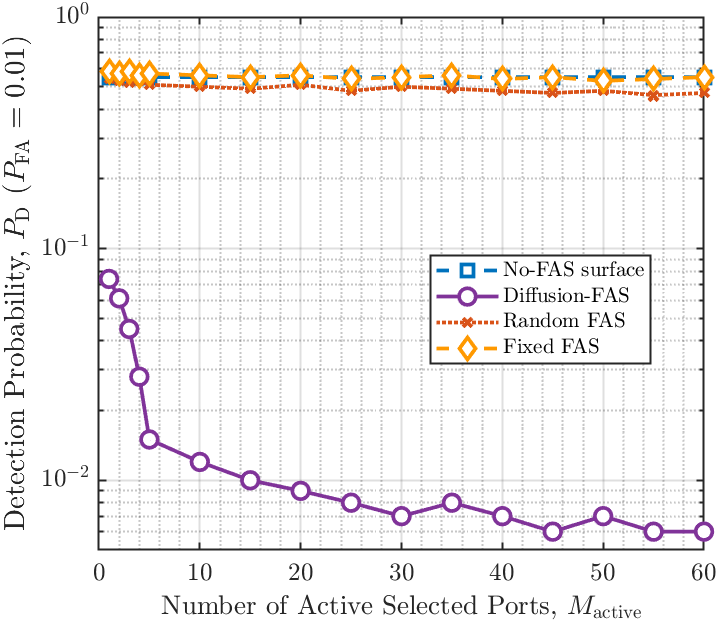}\label{fig:stealth_01}}
\subfigure[]{\includegraphics[width=0.8\columnwidth]{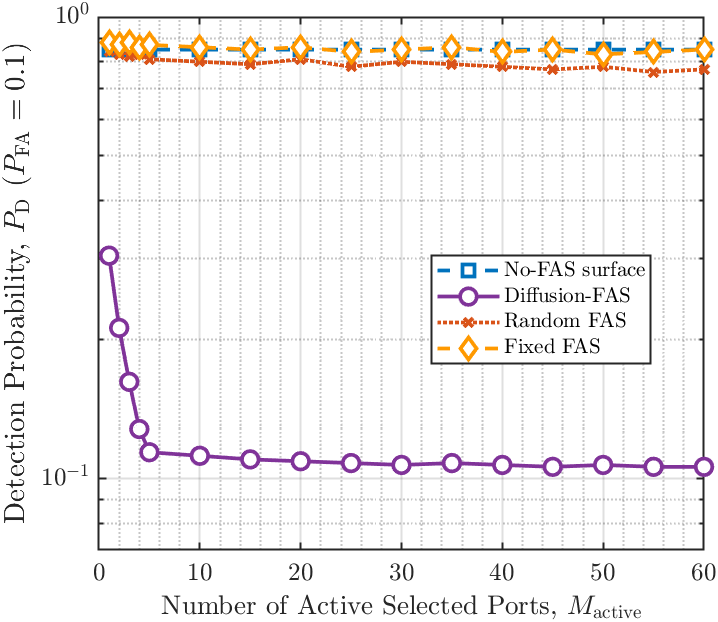}\label{fig:stealth_1}}
\caption{Probability of detection ($P_D$) vs. $M_{\mathrm{active}}$ for Case 1.}\label{fig:stealth_results}
\vspace{-5mm}
\end{figure}

The stealth performance for Case 1 is evaluated in Fig. \ref{fig:stealth_results}, which illustrates the probability of detection ($P_D$) as a function of the active port count ($M_{\mathrm{active}}$) for two different detection thresholds ($P_{\mathrm{FA}} = 0.01$ and $0.1$). A distinct optimization trend is observed for the proposed Diffusion-FAS policy in the low-$M_{\mathrm{active}}$ regime; as the number of selectable ports increases from $1$ to $5$, the generative model leverages the additional spatial DoFs to identify deeper localized fades, resulting in a sharp non-linear drop in $P_D$. While the static No-FAS surface and fixed-subset baselines remain highly visible, saturating near $P_D \approx 0.8$ under the relaxed $P_{\mathrm{FA}} = 0.1$ constraint, Diffusion-FAS maintains a consistent stealth floor nearly two orders of magnitude lower than the uninformed Random FAS selection. These results confirm that by intelligently navigating the environmental spatial correlation, FAS can effectively ``hide'' the user's signature from adversarial sensors.

\vspace{-2mm}
\section{Conclusion}\label{sec:conclusion}
This paper proposed a generative-AI-driven framework, termed Diffusion-FAS, which redefines the ISAC paradigm by treating sensing objectives as a dynamic spatial selection problem within FAS apertures. By leveraging the denoising trajectories of conditional diffusion models, our framework reconstructs the spatial fading manifold from sparse observations, enabling the real-time synthesis of port configurations optimized for sensing-centric spatial control. Our numerical results demonstrated that by intelligently navigating this spatial manifold, Diffusion-FAS can either suppress a user's sensing visibility by two orders of magnitude in generative stealth scenarios or perfectly isolate target reflections from spatially adjacent interference. These findings confirmed that FAS provides a robust pathway to transcend traditional waveform-based ISAC limits, offering a resilient, secure solution.

\vspace{-3mm}

\ifCLASSOPTIONcaptionsoff
  \newpage
\fi

\bibliographystyle{IEEEtran}
\section*{REFERENCES}
\def\refname{\vadjust{\vspace*{-1em}}}

\end{document}